\RequirePackage{fix-cm}
\documentclass[twocolumn,epjc3]{svjour3}
\smartqed  
\RequirePackage{graphicx}
\RequirePackage{mathptmx}      
\RequirePackage[english]{babel}
\RequirePackage{breqn}
\RequirePackage[numbers,sort&compress]{natbib}
\journalname{submitted to Eur. Phys. J. Plus}
\begin{document}

\title{A general analytical solution for the variance-to-mean Feynman-alpha formulas for a two-group two-point, a two-group one-point and a one-group two-point cases.
}


\author{Dina Chernikova\thanksref{e1,addr1}
        \and
        Wang Ziguan\thanksref{addr1} 
        \and
        Imre P\'{a}zsit\thanksref{addr1} 
        \and
        L\'en\'ard P\'al\thanksref{addr2} 
}

\thankstext{e1}{e-mail: dina@nephy.chalmers.se}


\institute{Chalmers University of Technology, Department of Applied Physics, Nuclear Engineering, \\Fysikg{\aa}rden 4, SE-412 96 G\"oteborg, Sweden \label{addr1}
           \and
           Centre for Energy Research, Hungarian Academy of Sciences, H-1525 Budapest 114, POB 49, Hungary \label{addr2}
}

\date{submitted to Eur. Phys. J. Plus}

\maketitle

\begin{abstract}
This paper presents a full derivation of the variance-to-mean or Feynman-alpha formula in a two energy group- and two spatial region-treatment. The derivation is based on the Chapman - Kolmogorov equation with the inclusion of all possible neutron reactions and passage intensities between the two regions. In addition, the two-group one-region and the two-region one-group Feynman-alpha formulas, treated earlier in the literature for special cases, are extended for further types and positions of detectors. We focus on the possibility of using these theories for accelerator-driven systems and applications in the safeguards domain, such as the differential self-interrogation method and the differential die-away method. This is due to the fact that the predictions from the models which are currently used do not fully describe all the effects in the heavily reflected fast or thermal systems. Therefore, in conclusion a comparative study of the two-group two-region, the two-group one-region, the one-group two-region and the one-group one-region Feynman-alpha models is discussed.
\end{abstract}

\section{Introduction}
\label{intro}
In detection statistics, the relation between the average number of counts during a detection time $t$, ${<N>}$, and the fluctuations around this value, expressed by the variance ${<N>^{2}}$-${<N^{2}>}$, i.e. the variance to mean ratio
\begin{dmath*}[style={\small}]
Q^{2} \sim \frac{<N^{2}>-<N>^{2}}{<N>}
\end{dmath*}
is often used to characterise the statistics of the particle field detected. For instance in the case of neutrons emitted from a radioactive source following a simple Poisson statistics, this ratio is obviously equal to unity. However, for a neutron chain in a multiplying medium, such as a subcritical reactor with a source or a fissile sample with an inherent neutron source due to spontaneous fission, the branching character represented by the fission process has the consequence that the individual detections will not be independent, rather positive correlations exist between them. Hence the variance to mean ratio is larger than unity, and the deviation from unity carries information on the medium in which the branching process (neutron multiplication) took place.

This fact was used by Feynman and de Hoffmann in 1944-1956 \cite{Fermi,Feynman,Hoffmann} for the derivation of a formula for a branching process where the variance to mean was above unity, $Q{^{2}}=1+Y(t)$. The $Y(t)$-function became called the Feynman Y-function, characterising the deviation of the relative variance from unity. Both its time dependence, expressed by the prompt neutron decay constant $\alpha$, as well as its asymptotic value, carry information on the sought parameters of the system. The original application of these studies was related to the theoretical description of statistical fluctuations of the number of neutrons in multiplying medium or in other words, to the determination of the level of subcriticality. Therefore, the above-mentioned research remained classified for several years.

The fundamental principles of the Feynman-alpha theory have been extensively described in a number of publications, e.g. \cite{Imre}. About a decade ago the interest to this subject returned again in connection to the on-line measurement of subcritical reactivity of Accelerator-Driven Systems (ADS). Whereas the original Feynman-alpha formulas referred to a homogeneous system in an monoenergetic ("one-group") description \cite{Feynman}, dealing only with one exponent or decay constant, the further experiments, e.g. the Yalina \cite{Kiyavitskaya,Gohar,Talamo} and MUSE \cite{Soule}, showed the appearance of more than one decay constant and, therefore, the possible need of extension of the one-group one-region (also referred to as "one-point") Feynman-alpha formulas to more energy groups and spatial regions. Several attempts were made towards the explanation of multiple exponential modes by the spatial effects \cite{Munoz,Berglof}. By that time it was decided that the future ADS-systems will be driven by  pulsed neutron and spallation sources which lead to the extension of the theory of variance-to-mean formulas for a continuous source with Poisson statistics to the cases of pulsed and spallation neutron sources with different definition of the pulse shapes and pulsing manner \cite{Rana2009,Rana2011,Degweker2000,Yamane2000,Yamane2001,Yamane2002,Kitamura2003,Kitamura2004,Kitamura20044,Kitamura2005,Munoz2001,Degweker2003,Ceder2003,Imre2004}.
Latter analysis \cite{Croft} showed the close link between the application of Feynman-alpha formulas to subcriticality measurements and Safeguards.

In line with the above, the suggestion of new Safeguards technique for MOX/spent fuel assay \cite{Menlove}, the Differential Die-away Self-Interrogation (DDSI) technique, displayed the interest towards the energy-dependent aspects of neutron counting. In connection with this, the two-group Feynman-alpha theory was elaborated in \cite{Anderson2012} where delayed neutrons were neglected, and in \cite{Pal2012} with inclusion of delayed neutron precursors. However, fast fission and thermal detections were neglected in both papers. The results of further considerations of the importance of the energy-aspect in evaluation of the real systems shows that "a measured variance-to-mean ratio in fast systems may be contaminated by the energy-higher order mode effect except when the system is near-critical \cite{Yamamoto2013}".

In the light of recent advances in detector technologies in Safeguards towards the development of fast neutron detection systems with scintillators, the knowledge of the energy-dependent behavior of neutron counting became a very important issue to be taken into account in Feynman-alpha theory. The authors of \cite{Chapline2011} showed that the short and long time behavior of the $Y$-function can be used to assay the amount of $^{240}$Pu and the absolute amount of $^{239}$Pu+$^{241}$Pu in the reprocessed fuel. Therefore, one part of this paper is devoted to the derivation of the general case of one-point two-group Feynman-alpha formulas, when fast fission and thermal detections and delayed neutrons are included.
However in some cases, for example, when the fission chambers are used as  detectors, the energy importance makes way for the region-dependent aspect. This issue has not well been studied previously, although some expressions for the one-group two-region Feynman-alpha formulas can be found in \cite{Anderson20121}. However, even these investigations are limited to the case of delayed neutron precursors having been neglected and detections accounted for only in one region. Thus, the second part of this paper is devoted to the derivation of the general case of the two-point one-group Feynman-alpha formulas, when detections and delayed neutrons are accounted for in both regions.

It has to be noted that the present paper does not carry out fully an analysis of the diagnostic value of the obtained formulas the same way as it was made in the traditional works based on a one-group treatment in a single (infinite) homogeneous medium. In the traditional case the time dependence of the Feynman $Y(t)$ function is characterised essentially with one decay constant which can clearly be related to the subcriticality of the system. In the case of using two energy groups and two spatial regions, the number of decay constants increases and each of them becomes a much more involved function of the increased number of material properties (reaction intensities) that the treatment of different regions and energy intervals incurs. The sought system parameters become very involved functions of these decay constants, and no attempt is made in this paper on the investigation of how these parameters can be extracted from the measurements. This is deferred to later work. The objective of the present work is to give a clear and transparent derivation of
the various variance-to-mean formulas as functions of the reaction and transition intensities, and to compare the solutions for the different cases.

\section{The main concept and assumptions}
\label{sec:1}
In this paper, the two-point two-group, the two-group one-point (with delay neutrons) and the one-group two-point (with delayed neutrons) Feynman-alpha formulas were derived by using the Kolmogorov forward approach \cite{Imre}. In the general model used for derivations we assume that the neutron population consists of two groups of neutrons: fast (denoted as 1) and thermal (denoted as 2). Fast and thermal neutrons can undergo different reactions (i) listed below:
\begin{itemize}
  \item absorption (${i=a}$),
  \item fission (${i=f}$),
  \item detection (${i=d}$),
  \item removal from the fast group to the thermal (${i=r}$).
\end{itemize}
Unlike in the terminology, used in the traditional one-group treatments, absorbtion here stands only for capture. The decay constant of the delayed neutron precursors is given as $\lambda$. In addition, both the fast and thermal neutrons can transit from one region to the others, in both directions. In all models the source is considered as releasing n particles with probability \emph{${p_{q}(n)}$} at an emission event. In this paper a term \emph{"two-point"} has the same meaning as \emph{"two-region"}.
\subsection{The two-group two-point model}
\label{sec:2}
\begin{figure}[ht!]
\centering
\includegraphics[width=0.5\textwidth]{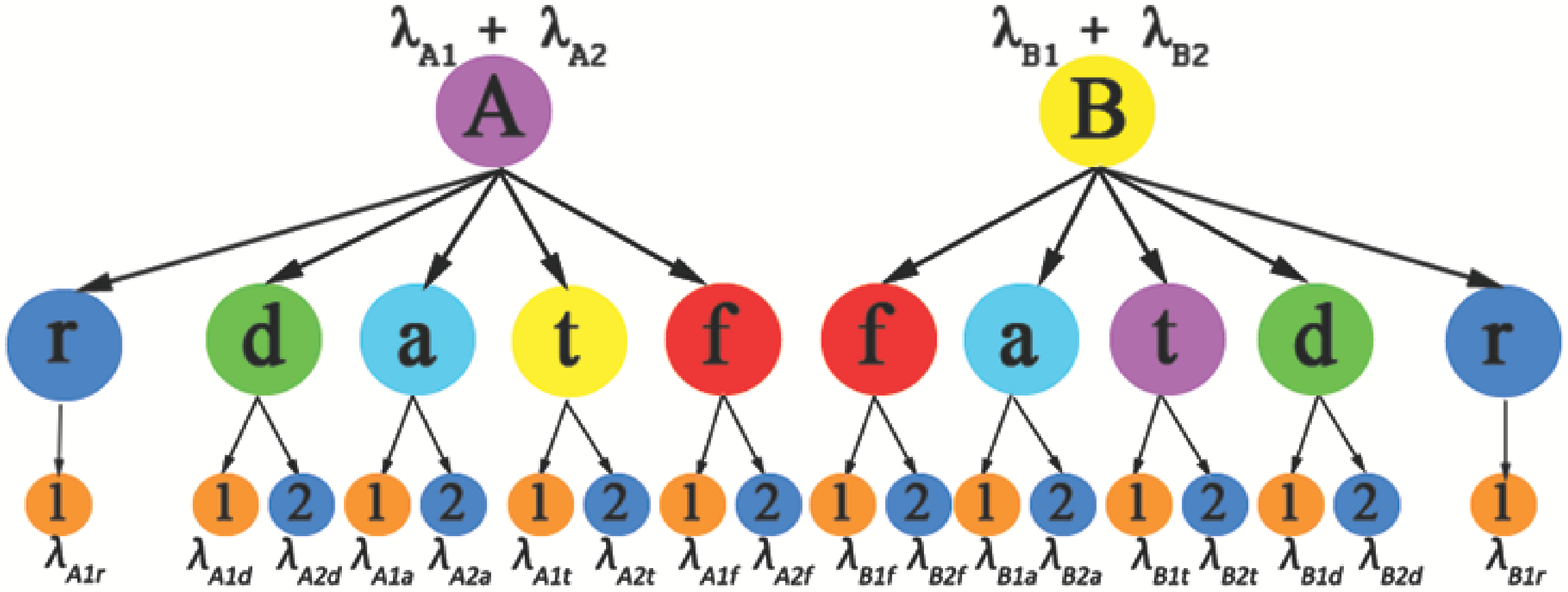}
\caption{A two-point two-group model of various processes which particles can undergo.}
\label{fig:1}
\end{figure}

For the two-group two-point model it was assumed that two adjacent infinite and homogeneous half-space regions (denoted as A and B) with different independent reaction intensities for absorption of fast and thermal neutrons (${\lambda_{A1a}}$, ${\lambda_{A2a}}$, ${\lambda_{B1a}}$, ${\lambda_{B2a}}$), fission induced by fast and thermal neutrons (${\lambda_{A1f}}$, ${\lambda_{A2f}}$, ${\lambda_{B1f}}$, ${\lambda_{B2f}}$) and detection of fast and thermal neutrons (${\lambda_{A1d}}$, ${\lambda_{B1d}}$, ${\lambda_{A2d}}$, ${\lambda_{B2d}}$). The two regions are coupled by two passage intensities (${\lambda_{A1t}}$, ${\lambda_{A2t}}$, ${\lambda_{B1t}}$, ${\lambda_{B2t}}$) in two different directions\footnote{${\lambda_{Ait}}$ describes the intensity of particles (group i) leaving region A for region B and ${\lambda_{Bit}}$ is the intensity of particles (group i) transferring to region A from region B.}. Thus, each of the reactions for the different groups of neutrons can be described by transition intensities, as shown in Figure \ref{fig:1}.
Total intensities including both the reactions and transitions between the regions for the fast and the thermal neutrons are denoted as ${\lambda_{A1}}$ and ${\lambda_{A2}}$, ${\lambda_{B1}}$ and ${\lambda_{B2}}$ for regions A and B, respectively:
\begin{dgroup*}
\begin{dmath*}[style={\small}]
\lambda_{A1} = \lambda_{A1a} +\lambda_{A1f} + \lambda_{A1t} + \lambda_{A1r} + \lambda_{A1d}
\end{dmath*}
\begin{dmath*}[style={\small}]
\lambda_{A2} = \lambda_{A2a} + \lambda_{A2f} + \lambda_{A2t} + \lambda_{A2d}
\end{dmath*}
\begin{dmath*}[style={\small}]
\lambda_{B1} = \lambda_{B1a}+ \lambda_{B1f} + \lambda_{B1t} + \lambda_{B1r} + \lambda_{B1d}
\end{dmath*}
\begin{dmath*}[style={\small}]
\lambda_{B2}=\lambda_{B2a} + \lambda_{B2f} + \lambda_{B2t} + \lambda_{B2d}
\end{dmath*}
\end{dgroup*}
The slowing down process, i.e. the removal of neutrons from the fast group to the thermal group is described by the removal reaction intensity $\lambda_{i=r=R}$. In the two-point two-group model we also include two extraneous compound Poisson sources of fast neutrons placed in different regions, A and/or B, with intensities ${S_A}$ and ${S_B}$. In the following, two special cases of the above general form will be described briefly. Because in the lower dimensionality of the special cases, inclusion of delayed neutrons is possible.
\subsection{The two-group one-point model (with delay neutrons)}
\label{sec:3}
\begin{figure}[ht!]
\centering
\includegraphics[width=0.33\textwidth]{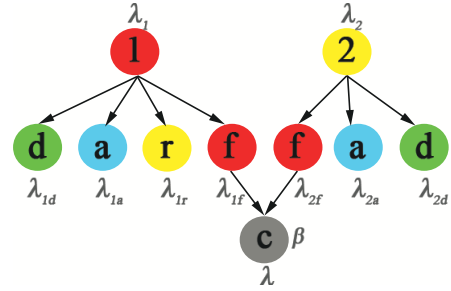}
\caption{A two-group one-point model of various processes which particles can undergo.}
\label{fig:25}
\end{figure}

In the two-group one-point Feynman-alpha model (Figure \ref{fig:25}), we assume that the medium is infinite and homogeneous. The neutron population consists of two groups of neutrons, fast and thermal. A compound Poisson source of fast neutrons with emission intensity ${S_1}$ is included in the model.
Thus, the total transition intensities for the fast and thermal neutrons, denoted as $\lambda_{1}$ and $\lambda_{2}$, are given as:
\begin{dgroup*}
\begin{dmath*}[style={\small}]
\lambda_1 = \lambda_{1a} + \lambda_{1f} + \lambda_{R} + \lambda_{1d}
\end{dmath*}
\begin{dmath*}[style={\small}]
\lambda_2 = \lambda_{2a} + \lambda_{2f} + \lambda_{2d}
\end{dmath*}
\end{dgroup*}
\subsection{The one-group two-point model (with delay neutrons)}
\label{sec:4}
\begin{figure}[ht!]
\centering
\includegraphics[width=0.4\textwidth]{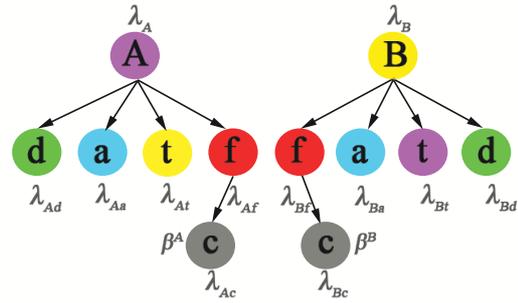}
\caption{A two-point one-group model of various processes which particles can undergo.}
\label{fig:3}
\end{figure}

The assumption behind the one-group two-point model is that the two adjacent homogeneous half-space regions (denoted as A and B) with independent reaction intensities for detection ($\lambda_{Ad}$, $\lambda_{Bd}$), absorption $\lambda_{Aa}$ and $\lambda_{Ba}$, and fission $\lambda_{Af}$ and $\lambda_{Bf}$ are coupled by two passage intensities $\lambda_{At}$ and $\lambda_{Bt}$ in two different directions. The decay constants of delayed neutron precursors are given as $\lambda_{Ac}$ and $\lambda_{Bc}$ for regions A and B, as shown in Figure \ref{fig:3}.
Thus, total transition intensities for region A and region B are denoted as $\lambda_{A}$ and $\lambda_{B}$:
\begin{dgroup*}
\begin{dmath*}[style={\small}]
\lambda_A = \lambda_{Aa} + \lambda_{Af} + \lambda_{At} + \lambda_{Ad}
\end{dmath*}
\begin{dmath*}[style={\small}]
\lambda_B = \lambda_{Ba} + \lambda_{Bf} + \lambda_{Bt} + \lambda_{Bd}
\end{dmath*}
\end{dgroup*}
In the model we include two compound Poisson sources of fast neutrons in regions A and B with emission intensities $S_A$ and $S_B$, respectively. The sources are considered as releasing $n$ particles in one emission with the probability distributions of $p_A(n)$ and $p_B(n)$, respectively. For the induced fission reaction, we consider that $k$ neutrons and $l$ delayed neutron precursors are emitted with the probability distributions $f_{A} (k,l)$ and $f_{B} (k,l)$ for the fission reaction in region A and region B, respectively.
\section{Theoretical formulas}
\label{sec:5}
Based on the main concept and assumptions used the two-group two-region, the two-group one-region and the two-region one-group Feynman-alpha formulas are elaborated as below.
\subsection{Two-point two-group Feynman-alpha theory}
\label{sec:8}
In order to derive the two-point two-group Feynman-alpha theory let us assume that the source ${S_{A}/S_{B}}$ is switched on in the region A/B at the time ${t_{0} \leq t}$, while the detection process is started at the fixed time instant ${t_{d}}$, where ${t_{d} \leq t}$ and ${t_{d} \geq t_{0}}$. Let the random processes $N_{A1}(t)$, $N_{B1}(t)$, $N_{A2}(t)$ and $N_{B2}(t)$ represent the number of fast neutrons in region A, fast neutrons in region B, thermal neutrons in region A and thermal neutrons in region B at the time ${t \geq 0}$ and ${Z_{A1}(t,t_{d})}$, ${Z_{A2}(t,t_{d})}$, ${Z_{B1}(t,t_{d})}$, ${Z_{B2}(t,t_{d})}$ - the number of fast and thermal particle detections in the regions A and B in the time interval [${t_{d}}$, ${t}$], respectively. For convenience, we consider \emph{${t_{d}}$=0}. Thus, the joint probability of having $N_{A1}$ fast neutrons in region A, $N_{B1}$ fast neutrons in region B, $N_{A2}$ thermal neutrons in region A, $N_{B2}$ thermal neutrons in region B at time $t$, $Z_{A1}$ fast neutrons have been detected in region A, and $Z_{B1}$ fast neutrons have been detected in region B, $Z_{A2}$ thermal neutrons have been detected in region A, and $Z_{B2}$ thermal neutrons have been detected in region B during the period of time ${t-t_{d}\geq 0}$ can be defined as:

${P(N_{A1},N_{A2},N_{B1},N_{B2},Z_{A1},Z_{B1},Z_{A2},Z_{B2},t|t_0)}$. 

By summing up the probabilities of all mutually exclusive events of the particle not having or having a specific reaction within the infinitesimally small time interval d$t$, one can write:
\begin{dmath*}[style={\small}]
\frac{\partial P(N_{A1},N_{A2},N_{B1},N_{B2},Z_{A1},Z_{B1},Z_{A2},Z_{B2},t)}{\partial t}=
-(\lambda_{A1}N_{A1}+\lambda_{A2}N_{A2}+\lambda_{B1}N_{B1}+\lambda_{B2}N_{B2}+S_{A}+S_{B})P(N_{A1},N_{A2},N_{B1},N_{B2},Z_{A1},Z_{B1},Z_{A2},Z_{B2},t)
+\lambda_{A1a}(N_{A1}+1)P(N_{A1}+1,N_{A2},N_{B1},N_{B2},Z_{A1},Z_{B1},Z_{A2},Z_{B2},t)
+\lambda_{A2a}(N_{A2}+1)P(N_{A1},N_{A2}+1,N_{B1},N_{B2},Z_{A1},Z_{B1},Z_{A2},Z_{B2},t)
+\lambda_{B1a}(N_{B1}+1)P(N_{A1},N_{A2},N_{B1}+1,N_{B2},Z_{A1},Z_{B1},Z_{A2},Z_{B2},t)
+\lambda_{B2a}(N_{B2}+1)P(N_{A1},N_{A2},N_{B1},N_{B2}+1,Z_{A1},Z_{B1},Z_{A2},Z_{B2},t)
+\lambda_{A1f}\sum_k^{N_{A1}+1}(N_{A1}+1-k)f_{A1}(k)P(N_{A1}+1-k,N_{A2},N_{B1},N_{B2},Z_{A1},Z_{B1},Z_{A2},Z_{B2},t)
+\lambda_{B1f}\sum_k^{N_{B1}+1}(N_{B1}+1-k)f_{B1}(k)P(N_{A1},N_{A2},N_{B1}+1-k,N_{B2},Z_{A1},Z_{B1},Z_{A2},Z_{B2},t)
\end{dmath*}
\begin{dmath*}[style={\small}]
+\lambda_{A2f}\sum_k^{N_{A1}}(N_{A2}+1)f_{A2}(k)P(N_{A1}-k,N_{A2}+1,N_{B1},N_{B2},Z_{A1},Z_{B1},Z_{A2},Z_{B2},t)
+\lambda_{B2f}\sum_k^{N_{B1}}(N_{B2}+1)f_{B2}(k)P(N_{A1},N_{A2},N_{B1}-k,N_{B2}+1,Z_{A1},Z_{B1},Z_{A2},Z_{B2},t)
+\lambda_{A1t}(N_{A1}+1)P(N_{A1}+1,N_{A2},N_{B1}-1,N_{B2},Z_{A1},Z_{B1},Z_{A2},Z_{B2},t)
+\lambda_{B1t}(N_{B1}+1)P(N_{A1}-1,N_{A2},N_{B1}+1,N_{B2},Z_{A1},Z_{B1},Z_{A2},Z_{B2},t)
+\lambda_{A2t}(N_{A2}+1)P(N_{A1},N_{A2}+1,N_{B1},N_{B2}-1,Z_{A1},Z_{B1},Z_{A2},Z_{B2},t)
+\lambda_{B2t}(N_{B2}+1)P(N_{A1},N_{A2}-1,N_{B1},N_{B2}+1,Z_{A1},Z_{B1},Z_{A2},Z_{B2},t)
+\lambda_{Ar}(N_{A1}+1)P(N_{A1}+1,N_{A2}-1,N_{B1},N_{B2},Z_{A1},Z_{B1},Z_{A2},Z_{B2},t)
+\lambda_{Br}(N_{B1}+1)P(N_{A1},N_{A2},N_{B1}+1,N_{B2}-1,Z_{A1},Z_{B1},Z_{A2},Z_{B2},t)
+\lambda_{A1d}(N_{A1}+1)P(N_{A1}+1,N_{A2},N_{B1},N_{B2},Z_{A1}-1,Z_{B1},Z_{A2},Z_{B2},t)
+\lambda_{B1d}(N_{B1}+1)P(N_{A1},N_{A2},N_{B1}+1,N_{B2},Z_{A1},Z_{B1}-1,Z_{A2},Z_{B2},t)
+\lambda_{A2d}(N_{A2}+1)P(N_{A1},N_{A2}+1,N_{B1},N_{B2},Z_{A1},Z_{B1},Z_{A2}-1,Z_{B2},t)
+\lambda_{B2d}(N_{B2}+1)P(N_{A1},N_{A2},N_{B1},N_{B2}+1,Z_{A1},Z_{B1},Z_{A2},Z_{B2}-1,t)
+S_A\sum_n^{N_{A1}}p_A(n)P(N_{A1}-n,N_{A2},N_{B1},N_{B2},Z_{A1},Z_{B1},Z_{A2},Z_{B2},t)
+S_B\sum_n^{N_{B1}}p_B(n)P(N_{A1},N_{A2},N_{B1}-n,N_{B2},Z_{A1},Z_{B1},Z_{A2},Z_{B2},t)
\end{dmath*}
with initial conditions
\begin{dmath*}[style={\small}]
P(N_{A1},N_{A2},N_{B1},N_{B2},Z_{A1},Z_{B1},Z_{A2},Z_{B2},t=t_0\mid t_0) = \delta_{N_{A1},0} \delta_{N_{A2},0} \delta_{N_{B1},0} \delta_{N_{B2},0} \delta_{Z_{A1},0} \delta_{Z_{A2},0} \delta_{Z_{B1},0} \delta_{Z_{B2},0}
\end{dmath*}
and
\begin{dmath*}[style={\small}]
\sum_{N_{A1}} \sum_{N_{A2}} \sum_{N_{B1}} \sum_{N_{B2}} P(N_{A1},N_{A2},N_{B1},N_{B2},Z_{A1},Z_{B1},Z_{A2},Z_{B2},t=t_d\mid t_0) = \delta_{Z_{A1},0} \delta_{Z_{A2},0} \delta_{Z_{B1},0} \delta_{Z_{B2},0}
\end{dmath*}
and ${f_{i}(k)}$ is the number distribution of neutrons in a fission of type \emph{i}.

This equation can be solved by using the generating function technique in the way similar to as described in \cite{Imre}.
By defining the following generating function for the probability distribution $P(N_{A1},N_{A2},N_{B1},N_{B2},Z_{A1},Z_{B1},Z_{A2},Z_{B2},t)$:
\begin{dmath*}[style={\small}]
G(X_{A},Y_{A},X_{B},Y_{B},M,N,O,P,t)= \sum_{N_{A1}} \sum_{N_{A2}} \sum_{N_{B1}} \sum_{N_{B2}} \sum_{Z_{A1}} \sum_{Z_{A2}} \sum_{Z_{B1}} \sum_{Z_{B2}} X_{A}^{N_{A1}} Y_{A}^{N_{A2}} X_{B}^{N_{B1}} Y_{B}^{N_{B2}} M^{Z_{A1}} N^{Z_{A2}} O^{Z_{B1}} P^{Z_{B2}} * P(N_{A1},N_{A2},N_{B1},N_{B2},Z_{A1},Z_{B1},Z_{A2},Z_{B2},t)
\end{dmath*}
with initial condition for ${t_0 \leq 0}$
\begin{dmath*}[style={\small}]
G(X_{A},Y_{A},X_{B},Y_{B},M,N,O,P,t=t_0\mid t_0) = 1
\end{dmath*}
and
\begin{dmath*}[style={\small}]
G(1,1,1,1,M,N,O,P,t=t_d\mid t_0) = 1
\end{dmath*}
the following partial differential equation is obtained:
\begin{dmath*}[style={\small}]
\frac{\partial G}{\partial t}= [\lambda_{A1a} + \lambda_{Ar}Y_{A} + \lambda_{A1t}X_{B} + \lambda_{A1d}M - \lambda_{A1}X_{A} + q_{A1}(X_{A}) \lambda_{A1f}] \frac{\partial G}{\partial X_{A}} + [\lambda_{A2a} + \lambda_{A2t}Y_{B} + \lambda_{A2d}N - \lambda_{A2}Y_{A} + q_{A2}(X_{A}) \lambda_{A2f}] \frac{\partial G}{\partial Y_{A}} +  [\lambda_{B1a} + \lambda_{Br}Y_{B} + \lambda_{B1t}X_{A} + \lambda_{B1d}O - \lambda_{B1}X_{B} + q_{B1}(X_{B}) \lambda_{B1f}] \frac{\partial G}{\partial X_{B}} + [\lambda_{B2a} + \lambda_{B2t}Y_{A} + \lambda_{B2d}P - \lambda_{B2}Y_{B} + q_{B2}(X_{B}) \lambda_{B2f}] \frac{\partial G}{\partial Y_{B}} + S_{A} [r_{A}(X_{A}) - 1] G + S_{B} [r_{B}(X_{B}) - 1] G,
\end{dmath*}
where
\begin{dgroup*}
\begin{dmath*}[style={\small}]
q_i (X) = \sum_k X^k f_{if}(k)
\end{dmath*}
\begin{dmath*}[style={\small}]
r(X) = \sum_n p_q(n) X^n
\end{dmath*}
\end{dgroup*}
For the sake of simplicity, some identities are used in the solution as below ($i=1,2$):
\begin{dgroup*}
\begin{dmath*}[style={\small}]
\left.\frac{\partial}{\partial X}r (X)\right|_{X=1}= \sum_n n p_{q}(n) =r'
\end{dmath*}
\begin{dmath*}[style={\small}]
\left.\frac{\partial^2}{\partial X^2}r (X)\right|_{X=1}= \sum_n n(n-1) p_{q}(n) =r''
\end{dmath*}
\end{dgroup*}
Thus, ${{\nu_{Ai}}'}$ (${{q_{Ai}}'}$), ${{\nu_{Bi}}'}$ (${{q_{Bi}}'}$), ${{\nu_{Ai}}''}$ (${{q_{Ai}}''}$), ${{\nu_{Bi}}''}$ (${{q_{Bi}}''}$) and ${{r_{A}}'}$, ${{r_{B}}'}$, ${{r_{A}}''}$, ${{r_{B}}''}$ stand for the first and second factorial moments of the number of neutrons emitted in a fission process and in a source event, respectively. The index ${i=1,2}$ denotes fission induced by fast or thermal neutrons, respectively.
In a steady subcritical medium with a steady source, when $t_0 \to -\infty$, the following stationary solutions for the neutron populations ${\bar{N}_{A1}}$, ${\bar{N}_{A2}}$, ${\bar{N}_{B1}}$, ${\bar{N}_{B2}}$, ${\bar{Z}_{A1}}$, ${\bar{Z}_{B1}}$, ${\bar{Z}_{A2}}$, ${\bar{Z}_{B2}}$ are obtained as below:
\begin{dgroup*}
\begin{dmath*}[style={\small}]
\bar{N}_{A1} = \frac{1}{\omega _1 \omega _2 \omega _3 \omega _4}(S_B r_B' \left(\lambda _{{A2}} \lambda _{{B1t}} \lambda _{{B2}}+\lambda _{{B2t}} \left(-\lambda _{{A2t}} \lambda _{{B1t}}+\lambda _{{A2f}} \lambda _{{Br}} \nu_{{A2}}'\right)\right)+ +S_A r_A' \left(\lambda _{{A2t}} \lambda _{{B2t}} \left(-\lambda _{{B1}}+\lambda _{{B1f}} {\nu _{B1}}' \right)+\lambda _{{A2}} \left(\lambda _{{B1}} \lambda _{{B2}}-\lambda _{{B1f}} \lambda _{{B2}} {\nu _{B1}}'-\lambda _{{B2f}} \lambda _{{Br}} \nu _{{B2}}' \right)\right))
\end{dmath*}
\begin{dmath*}[style={\small}]
\bar{N}_{B1} = \frac{1}{\omega _1 \omega _2 \omega _3 \omega _4}(S_A r_A'\left(\lambda _{{A1t}} \left(\lambda _{{A2}} \lambda _{{B2}}-\lambda _{{A2t}} \lambda _{{B2t}}\right)+\lambda _{{A2t}} \lambda _{{Ar}} \lambda _{{B2f}} \nu _{{B2}}' \right) +S_B r_B' \left(\lambda _{{A1}} \left(\lambda _{{A2}} \lambda _{{B2}}-\lambda _{{A2t}} \lambda _{{B2t}}\right)+\lambda _{{A1f}} \left(-\lambda _{{A2}} \lambda _{{B2}}+\lambda _{{A2t}} \lambda _{{B2t}}\right){\nu _{A1}}'-\lambda _{{A2f}} \lambda _{{Ar}} \lambda _{{B2}} \nu _{{A2}}' \right))
\end{dmath*}
\begin{dmath*}[style={\small}]
\bar{N}_{A2}   =
\frac{1}{\omega _1 \omega _2 \omega _3 \omega _4}(S_A r_A'\left(\lambda _{{A1t}} \lambda _{{B2t}} \lambda _{{Br}}+\lambda _{{Ar}} \left(\lambda _{{B1}} \lambda _{{B2}}-\lambda _{{B1f}} \lambda _{{B2}} {\nu _{B1}}'-\lambda _{{B2f}} \lambda _{{Br}} \nu _{{B2}}' \right)\right) +S_B r_B' \left(\lambda _{{Ar}} \lambda _{{B1t}} \lambda _{{B2}}+\lambda _{{B2t}} \lambda _{{Br}} \left(\lambda _{{A1}}-\lambda _{{A1f}}{\nu _{A1}}'\right)\right))
\end{dmath*}
\begin{dmath*}[style={\small}]
 \bar{N}_{B2} =\frac{1}{\omega _1 \omega _2 \omega _3 \omega _4}(S_A r_A' \left(\lambda _{{A1t}} \lambda _{{A2}} \lambda _{{Br}}+\lambda _{{A2t}} \lambda _{{Ar}} \left(\lambda _{{B1}}-\lambda _{{B1f}} {\nu _{B1}}'\right)\right) +S_B r_B'\left(\lambda _{{A2t}} \lambda _{{Ar}} \lambda _{{B1t}}+\lambda _{{Br}} \left(\lambda _{{A1}} \lambda _{{A2}}-\lambda _{{A1f}} \lambda _{{A2}}{\nu _{A1}}'-\lambda _{{A2f}} \lambda _{{Ar}} \nu _{{A2}}'\right)\right))
 \end{dmath*}
\begin{dmath*}[style={\small}]
\bar{Z}_{A1} =\lambda_{A1d} \bar{N}_{A1} t
\end{dmath*}
\begin{dmath*}[style={\small}]
\bar{Z}_{B1} =\lambda_{B1d} \bar{N}_{B1} t
\end{dmath*}
\begin{dmath*}[style={\small}]
\bar{Z}_{A2}  =\lambda_{A2d} \bar{N}_{A2} t
\end{dmath*}
\begin{dmath*}[style={\small}]
\bar{Z}_{B2}=\lambda_{B2d} \bar{N}_{B2} t
\end{dmath*}
\end{dgroup*}
By introducing the modified second factorial moment of the random variables a and b as follows ${\mu_{aa}\equiv<a(a-1)>-<a>^{2}}$=${\sigma_{a}^{2}}$ - ${<a>}$,${\mu_{ab}\equiv<ab>-<a><b>}$ and then taking cross- and auto-derivatives, the following system of differential equations of modified second factorial moments (${\mu _{{X_A X_A}}}$, ${\mu _{{ X_B X_B}}}$, ${\mu _{{ X_A Y_A}}}$, ${\mu _{{X_A Y_B}}}$, ${\mu _{{ X_A X_B}}}$, ${\mu _{{ X_B Y_A}}}$, ${\mu _{{ X_B Y_B}}}$) for the neutron population are obtained as below:
\begin{dgroup*}
\begin{dmath*}[style={\small}]
\frac{\partial}{\partial t} \mu_{X_A X_A} = 2 \lambda _{{B1t}}  \mu_{X_A X_B}+ 2 \lambda _{{A2f}} {\nu_{A2}}'\mu_{X_A Y_A}+2(\lambda _{{A1f}}  {\nu_{A1}}'-\lambda _{{A1}})\mu_{X_A X_A} +S_A r_A''+\lambda _{{A2f}} {\nu_{A2}}'' \bar{N}_{A2}
+\lambda _{{A1f}} {\nu_{A1}}''\bar{N}_{A1}
\end{dmath*}
\begin{dmath*}[style={\small}]
\frac{\partial}{\partial t} \mu_{X_A Y_A} = \lambda _{{B1t}}  \mu_{Y_A X_B}+\lambda _{{A2f}}{\nu_{A2}}'  \mu_{Y_A Y_A}+\lambda _{{B2t}}  \mu_{X_A Y_B}+\left(\lambda _{{A1f}} \nu_{{A1}}' -\lambda _{{A1}}-\lambda _{{A2}} \right)  \mu_{X_A Y_A}+\lambda _{{Ar}} \mu_{X_A X_A}
\end{dmath*}
\begin{dmath*}[style={\small}]
\frac{\partial}{\partial t} \mu_{X_A X_B} = \lambda _{{B1t}}  \mu_{X_B X_B}+\lambda _{{A2f}}{\nu_{A2}}'   \mu_{Y_A X_B}+\lambda _{{B2f}} {\nu_{B2}}'  \mu_{X_A Y_B} +\left(\lambda _{{A1f}} \nu_{{A1}}' -\lambda _{{A1}}\right) \mu_{X_A X_B}+\left(\lambda _{{B1f}} {\nu_{B1}}'-\lambda _{{B1}}\right)  \mu_{X_A X_B}+\lambda _{{A1t}} \mu_{X_A X_A}
\end{dmath*}
\begin{dmath*}[style={\small}]
\frac{\partial}{\partial t} \mu_{X_A Y_B} =  \lambda _{{B1t}}  \mu_{X_B Y_B}+\lambda _{{A2f}}{\nu_{A2}}'  \mu_{Y_A Y_B}+\left(\lambda _{{A1f}} \nu_{{A1}}' -\lambda _{{A1}}-\lambda _{{B2}} \right) \mu_{X_A Y_B}+\lambda _{{Br}}  \mu_{X_A X_B}+\lambda _{{A2t}}  \mu_{X_A Y_A}
\end{dmath*}
\begin{dmath*}[style={\small}]
\frac{\partial}{\partial t} \mu_{X_B X_B} =  2\lambda _{{B2f}} {\nu_{B2}}'\mu_{X_B Y_B}+2(\lambda _{{B1f}} {\nu_{B1}}'-\lambda _{{B1}} )\mu_{X_B X_B}+2 \lambda _{{A1t}}  \mu_{X_A X_B} +S_B r_B''+\lambda _{{B2f}} {\nu_{B2}}''\bar{N}_{B2}+\lambda _{{B1f}} {\nu_{B1}}'' \bar{N}_{B1}
\end{dmath*}
\begin{dmath*}[style={\small}]
\frac{\partial}{\partial t} \mu_{Y_A X_B } =   \lambda _{{B2t}}  \mu_{X_B Y_B}+\lambda _{{B2f}} {\nu_{B2}}'  \mu_{Y_A Y_B}+\left(\lambda _{{B1f}} {\nu_{B1}}'-\lambda _{{B1}}-\lambda _{{A2}} \right) \mu_{Y_A X_B}+\lambda _{{Ar}}  \mu_{X_A X_B}+\lambda _{{A1t}}  \mu_{X_A Y_A}
\end{dmath*}
\begin{dmath*}[style={\small}]
\frac{\partial}{\partial t} \mu_{X_B Y_B} =  \lambda _{{B2f}} {\nu_{B2}}' \mu_{Y_B Y_B}+\left(\lambda _{{B1f}} {\nu_{B1}}'-\lambda _{{B1}}-\lambda _{{B2}} \right) \mu_{X_B Y_B}+\lambda _{{Br}}  \mu_{X_B X_B}+\lambda _{{A2t}} \mu_{Y_A X_B}+\lambda _{{A1t}}  \mu_{X_A Y_B}
\end{dmath*}
\begin{dmath*}[style={\small}]
\frac{\partial}{\partial t} \mu_{Y_A Y_A} =  2 \lambda _{{B2t}}  \mu_{Y_A Y_B}-2 \lambda _{{A2}} \mu_{Y_A Y_A}+2 \lambda _{{Ar}}  \mu_{X_A Y_A}
\end{dmath*}
\begin{dmath*}[style={\small}]
\frac{\partial}{\partial t} \mu_{Y_A Y_B} =  \lambda _{{B2t}}\mu_{Y_B Y_B}-\lambda _{{A2}} \mu_{Y_A Y_B}-\lambda _{{B2}}  \mu_{Y_A Y_B}+\lambda _{{Br}} \mu_{Y_A X_B}+\lambda _{{A2t}}  \mu_{Y_A Y_A}+\lambda _{{Ar}}  \mu_{X_A Y_B}
\end{dmath*}
\begin{dmath*}[style={\small}]
\frac{\partial}{\partial t} \mu_{Y_B Y_B} =  -2 \lambda _{{B2}}\mu_{Y_B Y_B}+2 \lambda _{{Br}} \mu_{X_B Y_B}+2 \lambda _{{A2t}}  \mu_{Y_A Y_B}
\end{dmath*}
\end{dgroup*}
This system can be solved in the stationary state (when left hand sides are equal to 0).
The final expression of two-point two-group Feynman-alpha formulas for fast detections is given as below:
\begin{dmath*}[style={\small}]
\frac{\sigma_{ZZ}^2(t)}{\bar{Z}_{A1/A2/B1/B2}} = 1+Y(t)=1 + \sum_{i=1}^{4} Y_i (1 - \frac{1-e^{- \omega_i t}}{\omega_i t})
\end{dmath*}
The four roots, namely ${\omega_1}$, ${\omega_2}$, ${\omega_3}$ and ${\omega_4}$ can be obtained by solving the forth order characteristic equation in ${\omega}$ with known coefficients a, b, c, d, obtained from the temporal Laplace transform of the time-dependent equations for ${\mu _{{Z_A Z_B}}}$ etc.:
\begin{dmath*}[style={\small}]
\omega^4+ a\cdot\omega^3+b\cdot\omega^2+c\cdot\omega+d = 0
\end{dmath*}
where
\begin{dgroup*}
\begin{dmath*}[style={\small}]
  a=  \lambda _{{A1}}+\lambda _{{A2}}+\lambda _{{B1}}+\lambda _{{B2}}-\lambda _{{A1f}}  {\nu_{A1}}'-\lambda _{{B1f}} {\nu_{B1}}'
\end{dmath*}
\begin{dmath*}[style={\small}]
 b= -\lambda _{{A1t}} \lambda _{{B1t}}+\lambda _{{A2}} \lambda _{{B2}}-\lambda _{{A2t}} \lambda _{{B2t}}-\lambda _{{A2}} \left(\lambda _{{A1f}}  {\nu_{A1}}'-\lambda _{{A1}}\right)-\lambda _{{B2}} \left(\lambda _{{A1f}}  {\nu_{A1}}'-\lambda _{{A1}}\right)-\lambda _{{A2}} \left(\lambda _{{B1f}} {\nu_{B1}}'-\lambda _{{B1}}\right)-\lambda _{{B2}} \left(\lambda _{{B1f}} {\nu_{B1}}'-\lambda _{{B1}}\right)+\left(\lambda _{{A1f}}  {\nu_{A1}}'-\lambda _{{A1}}\right) \left(\lambda _{{B1f}} {\nu_{B1}}'-\lambda _{{B1}}\right)-\lambda _{{A2f}} \lambda _{{Ar}} {\nu_{A2}}' -\lambda _{{B2f}} \lambda _{{Br}}  {\nu_{B2}}'
\end{dmath*}
\begin{dmath*}[style={\small}]
 c=  -\lambda _{{A1t}} \lambda _{{A2}} \lambda _{{B1t}}-\lambda _{{A1t}} \lambda _{{B1t}} \lambda _{{B2}}-\lambda _{{A2}} \lambda _{{B2}} \left(\lambda _{{A1f}}  {\nu_{A1}}'-\lambda _{{A1}}\right)+\lambda _{{A2t}} \lambda _{{B2t}} \left(\lambda _{{A1f}}  {\nu_{A1}}'-\lambda _{{A1}}\right)-\lambda _{{A2}} \lambda _{{B2}} \left(\lambda _{{B1f}} {\nu_{B1}}'-\lambda _{{B1}}\right)+\lambda _{{A2t}} \lambda _{{B2t}} \left(\lambda _{{B1f}} {\nu_{B1}}'-\lambda _{{B1}}\right)+\lambda _{{A2}} \left(\lambda _{{A1f}}  {\nu_{A1}}'-\lambda _{{A1}}\right) \left(\lambda _{{B1f}} {\nu_{B1}}'-\lambda _{{B1}}\right)+\lambda _{{B2}} \left(\lambda _{{A1f}}  {\nu_{A1}}'-\lambda _{{A1}}\right) \left(\lambda _{{B1f}} {\nu_{B1}}'-\lambda _{{B1}}\right)-\lambda _{{A2f}} \lambda _{{Ar}} \lambda _{{B2}} {\nu_{A2}}' +\lambda _{{A2f}} \lambda _{{Ar}} \left(\lambda _{{B1f}} {\nu_{B1}}'-\lambda _{{B1}}\right) {\nu_{A2}}'-\lambda _{{A2}} \lambda _{{B2f}} \lambda _{{Br}}  {\nu_{B2}}' +\lambda _{{B2f}} \lambda _{{Br}} \left(\lambda _{{A1f}}  {\nu_{A1}}'-\lambda _{{A1}}\right)  {\nu_{B2}}'
\end{dmath*}
\begin{dmath*}[style={\small}]
 d=  -\lambda _{{A1t}} \lambda _{{A2}} \lambda _{{B1t}} \lambda _{{B2}}+\lambda _{{A1t}} \lambda _{{A2t}} \lambda _{{B1t}} \lambda _{{B2t}}+\lambda _{{A2}} \lambda _{{B2}} \left(\lambda _{{A1f}}  {\nu_{A1}}'-\lambda _{{A1}}\right) \left(\lambda _{{B1f}} {\nu_{B1}}'-\lambda _{{B1}}\right)-\lambda _{{A2t}} \lambda _{{B2t}} \left(\lambda _{{A1f}}  {\nu_{A1}}'-\lambda _{{A1}}\right) \left(\lambda _{{B1f}} {\nu_{B1}}'-\lambda _{{B1}}\right)-\lambda _{{A1t}} \lambda _{{A2f}} \lambda _{{B2t}} \lambda _{{Br}} {\nu_{A2}}'+\lambda _{{A2f}} \lambda _{{Ar}} \lambda _{{B2}} \left(\lambda _{{B1f}} {\nu_{B1}}'-\lambda _{{B1}}\right) {\nu_{A2}}' -\lambda _{{A2t}} \lambda _{{Ar}} \lambda _{{B1t}} \lambda _{{B2f}} {\nu_{B2}}'+\lambda _{{A2}} \lambda _{{B2f}} \lambda _{{Br}} \left(\lambda _{{A1f}}  {\nu_{A1}}'-\lambda _{{A1}}\right)  {\nu_{B2}}' +\lambda _{{A2f}} \lambda _{{Ar}} \lambda _{{B2f}} \lambda _{{Br}} {\nu_{A2}}'   {\nu_{B2}}'=\omega _1 \omega _2 \omega _3 \omega _4
\end{dmath*}
\end{dgroup*}
If detection of fast neutrons is performed in region A, then the functions ${Y_1}$, ${Y_2}$, ${Y_3}$ and ${Y_4}$ should be used in the form:
\begin{dgroup*}
\begin{dmath*}[style={\small}]
-Y_1 = \frac{2 \lambda _{A1d} \left(K_0-\omega _1 \left(\omega _1 \left(K_3 \omega _1-K_2\right)+K_1\right)\right)}{\bar{N}_{A1} \omega _1 \left(\omega _1-\omega _2\right) \left(\omega _1-\omega _3\right) \left(\omega _1-\omega _4\right)}
\end{dmath*}
\begin{dmath*}[style={\small}]
-Y_2 = \frac{2 \lambda _{A1d} \left(K_0-\omega _2 \left(\omega _2 \left(K_3 \omega _2-K_2\right)+K_1\right)\right)}{\bar{N}_{A1} \omega _2 \left(\omega _2-\omega _1\right) \left(\omega _2-\omega _3\right) \left(\omega _2-\omega _4\right)}
\end{dmath*}
\begin{dmath*}[style={\small}]
- Y_3 = \frac{2 \lambda _{A1d} \left(K_0-\omega _3 \left(\omega _3 \left(K_3 \omega _3-K_2\right)+K_1\right)\right)}{\bar{N}_{A1} \omega _3 \left(\omega _3-\omega _1\right) \left(\omega _3-\omega _2\right) \left(\omega _3-\omega _4\right)}
\end{dmath*}
\begin{dmath*}[style={\small}]
 - Y_4 = \frac{2 \lambda _{A1d}\left(K_0-\omega _4 \left(\omega _4 \left(K_3 \omega _4-K_2\right)+K_1\right)\right)}{\bar{N}_{A1} \omega _4 \left(\omega _4-\omega _1\right) \left(\omega _4-\omega _2\right) \left(\omega _4-\omega _3\right)}
\end{dmath*}
\end{dgroup*}
where,
\begin{dgroup*}
\begin{dmath*}[style={\small}]
K_3 = \mu _{{X_A X_A}}
\end{dmath*}
\begin{dmath*}[style={\small}]
K_2=\lambda _{A2f} \mu _{{ X_A Y_A}} q_{A2}'\left(X_A\right)+\lambda _{A2} \mu _{{ X_A X_A}}-\lambda _{B1f} \mu _{{ X_A X_A}} q_{B1}'\left(X_B\right)+\lambda _{B1} \mu _{{ X_A X_A}}+\lambda _{B1t} \mu _{{ X_A X_B}}+\lambda _{B2} \mu _{{ X_A X_A}}
\end{dmath*}
\begin{dmath*}[style={\small}]
K_1 =\lambda _{A2f} \lambda _{B1f} \mu _{{ X_A Y_A}} q_{A2}'\left(X_A\right) q_{B1}'\left(X_B\right)+\lambda _{A2f} \lambda _{B1} \mu _{{ X_A Y_A}} q_{A2}'\left(X_A\right)+\lambda _{A2f} \lambda _{B2} \mu _{{ X_A Y_A}} q_{A2}'\left(X_A\right)+\lambda _{A2f} \lambda _{B2t} \mu _{{ X_A Y_B}} q_{A2}'\left(X_A\right)-\lambda _{A2} \lambda _{B1f} \mu _{{ X_A X_A}} q_{B1}'\left(X_B\right)+\lambda _{A2} \lambda _{B1} \mu _{{ X_A X_A}}+\lambda _{A2} \lambda _{B1t} \mu _{{ X_A X_B}}+\lambda _{A2} \lambda _{B2} \mu _{{ X_A X_A}}-\lambda _{A2t} \lambda _{B2t} \mu _{{ X_A X_A}}-\lambda _{B1f} \lambda _{B2} \mu _{{ X_A X_A}} q_{B1}'\left(X_B\right)+\lambda _{B1t} \lambda _{B2f} \mu _{{ X_A Y_B}} q_{B2}'\left(X_B\right)-\lambda _{B2f} \lambda _{Br} \mu _{{ X_A X_A}} q_{B2}'\left(X_B\right)+\lambda _{B1} \lambda _{B2} \mu _{{ X_A X_A}}+\lambda _{B1t} \lambda _{B2} \mu _{{ X_A X_B}}
\end{dmath*}
\begin{dmath*}[style={\small}]
K_0= \lambda _{A2f} \lambda _{B1f} \lambda _{B2} \mu _{{ X_A Y_A}} q_{A2}'\left(X_A\right) q_{B1}'\left(X_B\right)-\lambda _{A2f} \lambda _{B1f} \lambda _{B2t} \mu _{{ X_A Y_B}} q_{A2}'\left(X_A\right) q_{B1}'\left(X_B\right)-\lambda _{A2f} \lambda _{B2f} \lambda _{Br} \mu _{{ X_A Y_A}} q_{A2}'\left(X_A\right) q_{B2}'\left(X_B\right)+\lambda _{A2f} \lambda _{B1} \lambda _{B2} \mu _{{ X_A Y_A}} q_{A2}'\left(X_A\right)+\lambda _{A2f} \lambda _{B1} \lambda _{B2t} \mu _{{ X_A Y_B}} q_{A2}'\left(X_A\right)+\lambda _{A2f} \lambda _{B2t} \lambda _{Br} \mu _{{ X_A X_B}} q_{A2}'\left(X_A\right)-\lambda _{A2} \lambda _{B1f} \lambda _{B2} \mu _{{ X_A X_A}} q_{B1}'\left(X_B\right)+\lambda _{A2} \lambda _{B1t} \lambda _{B2f} \mu _{{ X_A Y_B}} q_{B2}'\left(X_B\right)-\lambda _{A2} \lambda _{B2f} \lambda _{Br} \mu _{{ X_A X_A}} q_{B2}'\left(X_B\right)+\lambda _{A2} \lambda _{B1} \lambda _{B2} \mu _{{ X_A X_A}}+\lambda _{A2} \lambda _{B1t} \lambda _{B2} \mu _{{ X_A X_B}}+\lambda _{A2t} \lambda _{B1f} \lambda _{B2t} \mu _{{ X_A X_A}} q_{B1}'\left(X_B\right)+\lambda _{A2t} \lambda _{B1t} \lambda _{B2f} \mu _{{ X_A Y_A}} q_{B2}'\left(X_B\right)-\lambda _{A2t} \lambda _{B1} \lambda _{B2t} \mu _{{ X_A X_A}}-\lambda _{A2t} \lambda _{B1t} \lambda _{B2t} \mu _{{ X_A X_B}}
\end{dmath*}
\end{dgroup*}
It can be shown that:
\begin{dmath*}[style={\small}]
Y_0 = Y_1+Y_2 + Y_3+ Y_4=\frac{2K_0\lambda _{{A1d}} }{\omega _1\omega _2\omega _3\omega _4\bar{N}_{A1}}
\end{dmath*}
If a thermal neutron detector is placed in Region A, then the following ${Y_1}$, ${Y_2}$, ${Y_3}$ and ${Y_4}$ functions are to be used:
\begin{dgroup*}
\begin{dmath*}[style={\small}]
-Y_1= \frac{2 \lambda _{A2d} \left(L_0-\omega _1 \left(\omega _1 \left(L_3 \omega _1-L_2\right)+L_1\right)\right)}{\omega _1 \left(\omega _1-\omega _2\right) \left(\omega _1-\omega _3\right) \left(\omega _1-\omega _4\right) \bar{N}_{A2}}
\end{dmath*}
\begin{dmath*}[style={\small}]
-Y_2 =\frac{2 \lambda _{A2d} \left(L_0-\omega _2 \left(\omega _2 \left(L_3 \omega _2-L_2\right)+L_1\right)\right)}{\omega _2 \left(\omega _2-\omega _1\right) \left(\omega _2-\omega _3\right) \left(\omega _2-\omega _4\right) \bar{N}_{A2}}
\end{dmath*}
\begin{dmath*}[style={\small}]
- Y_3 = \frac{2 \lambda _{A2d} \left(L_0-\omega _3 \left(\omega _3 \left(L_3 \omega _3-L_2\right)+L_1\right)\right)}{\omega _3 \left(\omega _3-\omega _1\right) \left(\omega _3-\omega _2\right) \left(\omega _3-\omega _4\right) \bar{N}_{A2}}
\end{dmath*}
\begin{dmath*}[style={\small}]
- Y_4 = \frac{2 \lambda _{A2d} \left(L_0-\omega _4 \left(\omega _4 \left(L_3 \omega _4-L_2\right)+L_1\right)\right)}{\omega _4 \left(\omega _4-\omega _1\right) \left(\omega _4-\omega _2\right) \left(\omega _4-\omega _3\right) \bar{N}_{A2}}
\end{dmath*}
\end{dgroup*}
where,
\begin{dgroup*}
\begin{dmath*}[style={\small}]
L_3 =\mu _{{ Y_A Y_A}}
\end{dmath*}
\begin{dmath*}[style={\small}]
L_2=-\lambda _{{A1f}} \mu _{{ Y_A Y_A}} q_{{A1}}'\left(X_A\right)+\lambda _{{A1}} \mu _{{ Y_A Y_A}}+\lambda _{{Ar}} \mu _{{ X_A Y_A}}-\lambda _{{B1f}} \mu _{{ Y_A Y_A}} q_{{B1}}'\left(X_B\right)+\lambda _{{B1}} \mu _{{ Y_A Y_A}}+\lambda _{{B2}} \mu _{{ Y_A Y_A}}+\lambda _{{B2t}} \mu _{{ Y_B Y_A}}
\end{dmath*}
\begin{dmath*}[style={\small}]
L_1 =  \lambda _{{A1f}} \lambda _{{B1f}} \mu _{{ Y_A Y_A}} q_{{A1}}'\left(X_A\right) q_{{B1}}'\left(X_B\right)-\lambda _{{A1f}} \lambda _{{B1}} \mu _{{ Y_A Y_A}} q_{{A1}}'\left(X_A\right)-\lambda _{{A1f}} \lambda _{{B2}} \mu _{{ Y_A Y_A}} q_{{A1}}'\left(X_A\right)-\lambda _{{A1f}} \lambda _{{B2t}} \mu _{{ Y_B Y_A}} q_{{A1}}'\left(X_A\right)-\lambda _{{A1}} \lambda _{{B1f}} \mu _{{ Y_A Y_A}} q_{{B1}}'\left(X_B\right)+\lambda _{{A1}} \lambda _{{B1}} \mu _{{ Y_A Y_A}}+\lambda _{{A1}} \lambda _{{B2}} \mu _{{ Y_A Y_A}}+\lambda _{{A1}} \lambda _{{B2t}} \mu _{{ Y_B Y_A}}-\lambda _{{A1t}} \lambda _{{B1t}} \mu _{{ Y_A Y_A}}-\lambda _{{Ar}} \lambda _{{B1f}} \mu _{{ X_A Y_A}} q_{{B1}}'\left(X_B\right)+\lambda _{{Ar}} \lambda _{{B1}} \mu _{{ X_A Y_A}}+\lambda _{{Ar}} \lambda _{{B1t}} \mu _{{ X_B Y_A}}+\lambda _{{Ar}} \lambda _{{B2}} \mu _{{ X_A Y_A}}-\lambda _{{B1f}} \lambda _{{B2}} \mu _{{ Y_A Y_A}} q_{{B1}}'\left(X_B\right)-\lambda _{{B1f}} \lambda _{{B2t}} \mu _{{ Y_B Y_A}} q_{{B1}}'\left(X_B\right)-\lambda _{{B2f}} \lambda _{{Br}} \mu _{{ Y_A Y_A}} q_{{B2}}'\left(X_B\right)+\lambda _{{B1}} \lambda _{{B2}} \mu _{{ Y_A Y_A}}+\lambda _{{B1}} \lambda _{{B2t}} \mu _{{ Y_B Y_A}}+\lambda _{{B2t}} \lambda _{{Br}} \mu _{{ X_B Y_A}}
\end{dmath*}
\begin{dmath*}[style={\small}]
L_0= \lambda _{{A1f}} \lambda _{{B1f}} \lambda _{{B2}} \mu _{{ Y_A Y_A}} q_{{A1}}'\left(X_A\right) q_{{B1}}'\left(X_B\right)+\lambda _{{A1f}} \lambda _{{B1f}} \lambda _{{B2t}} \mu _{{ Y_B Y_A}} q_{{A1}}'\left(X_A\right) q_{{B1}}'\left(X_B\right)+\lambda _{{A1f}} \lambda _{{B2f}} \lambda _{{Br}} \mu _{{ Y_A Y_A}} q_{{A1}}'\left(X_A\right) q_{{B2}}'\left(X_B\right)-\lambda _{{A1f}} \lambda _{{B1}} \lambda _{{B2}} \mu _{{ Y_A Y_A}} q_{{A1}}'\left(X_A\right)-\lambda _{{A1f}} \lambda _{{B1}} \lambda _{{B2t}} \mu _{{ Y_B Y_A}} q_{{A1}}'\left(X_A\right)-\lambda _{{A1f}} \lambda _{{B2t}} \lambda _{{Br}} \mu _{{ X_B Y_A}} q_{{A1}}'\left(X_A\right)-\lambda _{{A1}} \lambda _{{B1f}} \lambda _{{B2}} \mu _{{ Y_A Y_A}} q_{{B1}}'\left(X_B\right)-\lambda _{{A1}} \lambda _{{B1f}} \lambda _{{B2t}} \mu _{{ Y_B Y_A}} q_{{B1}}'\left(X_B\right)-\lambda _{{A1}} \lambda _{{B2f}} \lambda _{{Br}} \mu _{{ Y_A Y_A}} q_{{B2}}'\left(X_B\right)+\lambda _{{A1}} \lambda _{{B1}} \lambda _{{B2}} \mu _{{ Y_A Y_A}}+\lambda _{{A1}} \lambda _{{B1}} \lambda _{{B2t}} \mu _{{ Y_B Y_A}}+\lambda _{{A1}} \lambda _{{B2t}} \lambda _{{Br}} \mu _{{ X_B Y_A}}-\lambda _{{A1t}} \lambda _{{B1t}} \lambda _{{B2}} \mu _{{ Y_A Y_A}}-\lambda _{{A1t}} \lambda _{{B1t}} \lambda _{{B2t}} \mu _{{ Y_B Y_A}}+\lambda _{{A1t}} \lambda _{{B2t}} \lambda _{{Br}} \mu _{{ X_A Y_A}}-\lambda _{{Ar}} \lambda _{{B1f}} \lambda _{{B2}} \mu _{{ X_A Y_A}} q_{{B1}}'\left(X_B\right)+\lambda _{{Ar}} \lambda _{{B1t}} \lambda _{{B2f}} \mu _{{ Y_B Y_A}} q_{{B2}}'\left(X_B\right)-\lambda _{{Ar}} \lambda _{{B2f}} \lambda _{{Br}} \mu _{{ X_A Y_A}} q_{{B2}}'\left(X_B\right)+\lambda _{{Ar}} \lambda _{{B1}} \lambda _{{B2}} \mu _{{ X_A Y_A}}+\lambda _{{Ar}} \lambda _{{B1t}} \lambda _{{B2}} \mu _{{ X_B Y_A}}
\end{dmath*}
\end{dgroup*}
It can be shown that:
\begin{dmath*}[style={\small}]
Y_0=Y_1+Y_2+Y_3+Y_4=\frac{2 L_0 \lambda _{A2d}}{\omega _1 \omega _2 \omega _3 \omega _4 \bar{N}_{A2}}
\end{dmath*}
For the case when a fast neutron detector is placed in Region B, the following ${Y_1}$, ${Y_2}$, ${Y_3}$ and ${Y_4}$ functions should be used:
\begin{dgroup*}
\begin{dmath*}[style={\small}]
-Y_1= \frac{2 \lambda _{B1d} \left(M_0-\omega _1 \left(\omega _1 \left(M_3 \omega _1-M_2\right)+M_1\right)\right)}{ \bar{N}_{B1} \omega _1 \left(\omega _1-\omega _2\right) \left(\omega _1-\omega _3\right) \left(\omega _1-\omega _4\right)}
\end{dmath*}
\begin{dmath*}[style={\small}]
-Y_2 =\frac{2 \lambda _{B1d} \left(M_0-\omega _2 \left(\omega _2 \left(M_3 \omega _2-M_2\right)+M_1\right)\right)}{\bar{N}_{B1}\omega _2 \left(\omega _2-\omega _1\right) \left(\omega _2-\omega _3\right) \left(\omega _2-\omega _4\right)}
\end{dmath*}
\begin{dmath*}[style={\small}]
- Y_3 = \frac{2 \lambda _{B1d} \left(M_0-\omega _3 \left(\omega _3 \left(M_3 \omega _3-M_2\right)+M_1\right)\right)}{\bar{N}_{B1}\omega _3 \left(\omega _3-\omega _1\right) \left(\omega _3-\omega _2\right) \left(\omega _3-\omega _4\right)}
\end{dmath*}
\begin{dmath*}[style={\small}]
- Y_4 = \frac{2 \lambda _{B1d} \left(M_0-\omega _4 \left(\omega _4 \left(M_3 \omega _4-M_2\right)+M_1\right)\right)}{\bar{N}_{B1} \omega _4 \left(\omega _4-\omega _1\right) \left(\omega _4-\omega _2\right) \left(\omega _4-\omega _3\right)}
\end{dmath*}
\end{dgroup*}
where,
\begin{dgroup*}
\begin{dmath*}[style={\small}]
M_3 =\mu _{{X_B X_B}}
\end{dmath*}
\begin{dmath*}[style={\small}]
M_2=-\lambda _{{A1f}} \mu _{{X_B X_B}} q_{{A1}}'\left(X_A\right)+\lambda _{{A1}} \mu _{{X_B X_B}}+\lambda _{{A1t}} \mu _{{X_A X_B}}+\lambda _{{A2}} \mu _{{X_B X_B}}+\lambda _{{B2f}} \mu _{{X_B Y_B}} q_{{B2}}'\left(X_B\right)+\lambda _{{B2}} \mu _{{X_B X_B}}
\end{dmath*}
\begin{dmath*}[style={\small}]
M_1 = -\lambda _{{A1f}} \lambda _{{A2}} \mu _{{X_B X_B}} q_{{A1}}'\left(X_A\right)-\lambda _{{A1f}} \lambda _{{B2f}} \mu _{{X_B Y_B}} q_{{A1}}'\left(X_A\right) q_{{B2}}'\left(X_B\right)-\lambda _{{A1f}} \lambda _{{B2}} \mu _{{X_B X_B}} q_{{A1}}'\left(X_A\right)+\lambda _{{A1t}} \lambda _{{A2f}} \mu _{{X_B Y_A}} q_{{A2}}'\left(X_A\right)-\lambda _{{A2f}} \lambda _{{Ar}} \mu _{{X_B X_B}} q_{{A2}}'\left(X_A\right)+\lambda _{{A1}} \lambda _{{A2}} \mu _{{X_B X_B}}+\lambda _{{A1}} \lambda _{{B2f}} \mu _{{X_B Y_B}} q_{{B2}}'\left(X_B\right)+\lambda _{{A1}} \lambda _{{B2}} \mu _{{X_B X_B}}+\lambda _{{A1t}} \lambda _{{A2}} \mu _{{X_A X_B}}+\lambda _{{A1t}} \lambda _{{B2}} \mu _{{X_A X_B}}+\lambda _{{A2}} \lambda _{{B2f}} \mu _{{X_B Y_B}} q_{{B2}}'\left(X_B\right)+\lambda _{{A2}} \lambda _{{B2}} \mu _{{X_B X_B}}+\lambda _{{A2t}} \lambda _{{B2f}} \mu _{{X_B Y_A}} q_{{B2}}'\left(X_B\right)-\lambda _{{A2t}} \lambda _{B2t} \mu _{{X_B X_B}}
\end{dmath*}
\begin{dmath*}[style={\small}]
M_0=-\lambda _{{A1f}} \lambda _{{A2}} \lambda _{{B2f}} \mu _{{X_B Y_B}} q_{{A1}}'\left(X_A\right) q_{{B2}}'\left(X_B\right)-\lambda _{{A1f}} \lambda _{{A2}} \lambda _{{B2}} \mu _{{X_B X_B}} q_{{A1}}'\left(X_A\right)-\lambda _{{A1f}} \lambda _{{A2t}} \lambda _{{B2f}} \mu _{{X_B Y_A}} q_{{A1}}'\left(X_A\right) q_{{B2}}'\left(X_B\right)+\lambda _{{A1f}} \lambda _{{A2t}} \lambda _{B2t} \mu _{{X_B X_B}} q_{{A1}}'\left(X_A\right)+\lambda _{{A1t}} \lambda _{{A2f}} \lambda _{{B2}} \mu _{{X_B Y_A}} q_{{A2}}'\left(X_A\right)+\lambda _{{A1t}} \lambda _{{A2f}} \lambda _{B2t} \mu _{{X_B Y_B}} q_{{A2}}'\left(X_A\right)-\lambda _{{A2f}} \lambda _{{Ar}} \lambda _{{B2f}} \mu _{{X_B Y_B}} q_{{A2}}'\left(X_A\right) q_{{B2}}'\left(X_B\right)-\lambda _{{A2f}} \lambda _{{Ar}} \lambda _{{B2}} \mu _{{X_B X_B}} q_{{A2}}'\left(X_A\right)+\lambda _{{A1}} \lambda _{{A2}} \lambda _{{B2f}} \mu _{{X_B Y_B}} q_{{B2}}'\left(X_B\right)+\lambda _{{A1}} \lambda _{{A2}} \lambda _{{B2}} \mu _{{X_B X_B}}+\lambda _{{A1}} \lambda _{{A2t}} \lambda _{{B2f}} \mu _{{X_B Y_A}} q_{{B2}}'\left(X_B\right)-\lambda _{{A1}} \lambda _{{A2t}} \lambda _{B2t} \mu _{{X_B X_B}}+\lambda _{{A1t}} \lambda _{{A2}} \lambda _{{B2}} \mu _{{X_A X_B}}-\lambda _{{A1t}} \lambda _{{A2t}} \lambda _{B2t} \mu _{{X_A X_B}}+\lambda _{{A2t}} \lambda _{{Ar}} \lambda _{{B2f}} \mu _{{X_A X_B}} q_{{B2}}'\left(X_B\right)
\end{dmath*}
\end{dgroup*}
It can be shown that:
\begin{dmath*}[style={\small}]
Y_0=Y_1+Y_2+Y_3+Y_4=\frac{2 M_0 \lambda _{B1d}}{\bar{N}_{B1} \omega _1 \omega _2 \omega _3 \omega _4}
\end{dmath*}
If a thermal neutron detector is placed in Region B, the following ${Y_1}$, ${Y_2}$, ${Y_3}$ and ${Y_4}$ functions should be used:
\begin{dgroup*}
\begin{dmath*}[style={\small}]
-Y_1= \frac{2 \lambda _{B2d} \left(P_0-\omega _1 \left(\omega _1 \left(P_3 \omega _1-P_2\right)+P_1\right)\right)}{\omega _1 \left(\omega _1-\omega _2\right) \left(\omega _1-\omega _3\right) \left(\omega _1-\omega _4\right)\bar{N}_{B2}}
\end{dmath*}
\begin{dmath*}[style={\small}]
-Y_2 =\frac{2 \lambda _{B2d} \left(P_0-\omega _2 \left(\omega _2 \left(P_3 \omega _2-P_2\right)+P_1\right)\right)}{\omega _2 \left(\omega _2-\omega _1\right) \left(\omega _2-\omega _3\right) \left(\omega _2-\omega _4\right)\bar{N}_{B2}}
\end{dmath*}
\begin{dmath*}[style={\small}]
- Y_3 = \frac{2 \lambda _{B2d} \left(P_0-\omega _3 \left(\omega _3 \left(P_3 \omega _3-P_2\right)+P_1\right)\right)}{\omega _3 \left(\omega _3-\omega _1\right) \left(\omega _3-\omega _2\right) \left(\omega _3-\omega _4\right) \bar{N}_{B2}}
\end{dmath*}
\begin{dmath*}[style={\small}]
- Y_4 = \frac{2 \lambda _{B2d} \left(P_0-\omega _4 \left(\omega _4 \left(P_3 \omega _4-P_2\right)+P_1\right)\right)}{\omega _4 \left(\omega _4-\omega _1\right) \left(\omega _4-\omega _2\right) \left(\omega _4-\omega _3\right)\bar{N}_{B2}}
\end{dmath*}
\end{dgroup*}
where,
\begin{dgroup*}
\begin{dmath*}[style={\small}]
P_3 =\mu _{{Y_B Y_B}}
\end{dmath*}
\begin{dmath*}[style={\small}]
P_2=-\lambda _{{A1f}} \mu _{{Y_B Y_B}} q_{{A1}}'\left(X_A\right)+\lambda _{{A1}} \mu _{{Y_B Y_B}}+\lambda _{{A2}} \mu _{{Y_B Y_B}}+\lambda _{{A2t}} \mu _{{Y_B Y_A}}-\lambda _{{B1f}} \mu _{{Y_B Y_B}} q_{{B1}}'\left(X_B\right)+\lambda _{{B1}} \mu _{{Y_B Y_B}}+\lambda _{{Br}} \mu _{{X_B Y_B}}
\end{dmath*}
\begin{dmath*}[style={\small}]
P_1 = -\lambda _{{A1f}} \lambda _{{A2}} \mu _{{Y_B Y_B}} q_{{A1}}'\left(X_A\right)-\lambda _{{A1f}} \lambda _{{A2t}} \mu _{{Y_B Y_A}} q_{{A1}}'\left(X_A\right)+\lambda _{{A1f}} \lambda _{{B1f}} \mu _{{Y_B Y_B}} q_{{A1}}'\left(X_A\right) q_{{B1}}'\left(X_B\right)-\lambda _{{A1f}} \lambda _{{B1}} \mu _{{Y_B Y_B}} q_{{A1}}'\left(X_A\right)-\lambda _{{A1f}} \lambda _{{Br}} \mu _{{X_B Y_B}} q_{{A1}}'\left(X_A\right)-\lambda _{{A2f}} \lambda _{{Ar}} \mu _{{Y_B Y_B}} q_{{A2}}'\left(X_A\right)+\lambda _{{A1}} \lambda _{{A2}} \mu _{{Y_B Y_B}}+\lambda _{{A1}} \lambda _{{A2t}} \mu _{{Y_B Y_A}}-\lambda _{{A1}} \lambda _{{B1f}} \mu _{{Y_B Y_B}} q_{{B1}}'\left(X_B\right)+\lambda _{{A1}} \lambda _{{B1}} \mu _{{Y_B Y_B}}+\lambda _{{A1}} \lambda _{{Br}} \mu _{{X_B Y_B}}-\lambda _{{A1t}} \lambda _{{B1t}} \mu _{{Y_B Y_B}}+\lambda _{{A1t}} \lambda _{{Br}} \mu _{{X_A Y_B}}-\lambda _{{A2}} \lambda _{{B1f}} \mu _{{Y_B Y_B}} q_{{B1}}'\left(X_B\right)+\lambda _{{A2}} \lambda _{{B1}} \mu _{{Y_B Y_B}}+\lambda _{{A2}} \lambda _{{Br}} \mu _{{X_B Y_B}}+\lambda _{{A2t}} \lambda _{{Ar}} \mu _{{X_A Y_B}}-\lambda _{{A2t}} \lambda _{{B1f}} \mu _{{Y_B Y_A}} q_{{B1}}'\left(X_B\right)+\lambda _{{A2t}} \lambda _{{B1}} \mu _{{Y_B Y_A}}
\end{dmath*}
\begin{dmath*}[style={\small}]
P_0=\lambda _{{A1f}} \lambda _{{A2}} \lambda _{{B1f}} \mu _{{Y_B Y_B}} q_{{A1}}'\left(X_A\right) q_{{B1}}'\left(X_B\right)-\lambda _{{A1f}} \lambda _{{A2}} \lambda _{{B1}} \mu _{{Y_B Y_B}} q_{{A1}}'\left(X_A\right)-\lambda _{{A1f}} \lambda _{{A2}} \lambda _{{Br}} \mu _{{X_B Y_B}} q_{{A1}}'\left(X_A\right)+\lambda _{{A1f}} \lambda _{{A2t}} \lambda _{{B1f}} \mu _{{Y_B Y_A}} q_{{A1}}'\left(X_A\right) q_{{B1}}'\left(X_B\right)-\lambda _{{A1f}} \lambda _{{A2t}} \lambda _{{B1}} \mu _{{Y_B Y_A}} q_{{A1}}'\left(X_A\right)+\lambda _{{A1t}} \lambda _{{A2f}} \lambda _{{Br}} \mu _{{Y_B Y_A}} q_{{A2}}'\left(X_A\right)+\lambda _{{A2f}} \lambda _{{Ar}} \lambda _{{B1f}} \mu _{{Y_B Y_B}} q_{{A2}}'\left(X_A\right) q_{{B1}}'\left(X_B\right)-\lambda _{{A2f}} \lambda _{{Ar}} \lambda _{{B1}} \mu _{{Y_B Y_B}} q_{{A2}}'\left(X_A\right)-\lambda _{{A2f}} \lambda _{{Ar}} \lambda _{{Br}} \mu _{{X_B Y_B}} q_{{A2}}'\left(X_A\right)-\lambda _{{A1}} \lambda _{{A2}} \lambda _{{B1f}} \mu _{{Y_B Y_B}} q_{{B1}}'\left(X_B\right)+\lambda _{{A1}} \lambda _{{A2}} \lambda _{{B1}} \mu _{{Y_B Y_B}}+\lambda _{{A1}} \lambda _{{A2}} \lambda _{{Br}} \mu _{{X_B Y_B}}-\lambda _{{A1}} \lambda _{{A2t}} \lambda _{{B1f}} \mu _{{Y_B Y_A}} q_{{B1}}'\left(X_B\right)+\lambda _{{A1}} \lambda _{{A2t}} \lambda _{{B1}} \mu _{{Y_B Y_A}}-\lambda _{{A1t}} \lambda _{{A2}} \lambda _{{B1t}} \mu _{{Y_B Y_B}}+\lambda _{{A1t}} \lambda _{{A2}} \lambda _{{Br}} \mu _{{X_A Y_B}}-\lambda _{{A1t}} \lambda _{{A2t}} \lambda _{{B1t}} \mu _{{Y_B Y_A}}-\lambda _{{A2t}} \lambda _{{Ar}} \lambda _{{B1f}} \mu _{{X_A Y_B}} q_{{B1}}'\left(X_B\right)+\lambda _{{A2t}} \lambda _{{Ar}} \lambda _{{B1}} \mu _{{X_A Y_B}}+\lambda _{{A2t}} \lambda _{{Ar}} \lambda _{{B1t}} \mu _{{X_B Y_B}}
\end{dmath*}
\end{dgroup*}
It can be shown that:
\begin{dmath*}[style={\small}]
Y_0=Y_1+Y_2+Y_3+Y_4=\frac{2 P_0 \lambda _{B2d}}{\bar{N}_{B2} \omega _1 \omega _2 \omega _3 \omega _4}.
\end{dmath*}
Quantitative examples of the Feynman Y(t) function will be given shortly.
\subsection{Two-group one-point Feynman-alpha theory (with delayed neutrons)}
\label{sec:6}
In order to derive the two-group one-point Feynman-alpha theory let us assume that the source S is switched on at the time ${t_{0} \leq t}$, while the detection process is started at the fixed time instant ${t_{d}}$, where ${t_{d} \leq t}$ and ${t_{d} \geq t_{0}}$. For convenience, we consider \emph{${t_{d}}$=0}. Let the random processes ${N_{1}(t)}$, ${N_{2}(t)}$ and ${C(t)}$ represent the number of fast neutrons, thermal neutrons and delayed neutron precursors at the time ${t \geq 0}$, and ${Z_{1}(t,t_{d})}$, ${Z_{2}(t,t_{d})}$ - the number of fast and thermal particle detections in the time interval \emph{[${t_{d}}$, t]}, respectively. Thus, the joint probability of having $N_{1}$ fast neutrons, $N_{2}$ thermal neutrons and C delayed neutron precursors present in system at time $t$, and that ${Z_{1}}$ fast neutrons and ${Z_{2}}$ thermal neutrons have been detected during the period ${t-t_{d}\geq 0}$ can be defined as ${P(N_1,N_2,C,Z_1,Z_2,t|t_0)}$.
By summing up the probabilities of the mutually exclusive events of the particle not having or having a specific reaction or that there is a source emission within the infinitesimally small time interval d$t$, one can write:
\begin{dmath*}[style={\small}]
\frac{\partial P(N_1,N_2,C,Z_1,Z_2,t)}{\partial t}
\end{dmath*}
\begin{dmath*}[style={\small}]
=-(\lambda_1N_1+\lambda_2N_2+\lambda C+S_1)P(N_1,N_2,C,Z_1,Z_2,t)
+\lambda_{1a}(N_1+1)P(N_1+1,N_2,C,Z_1,Z_2,t)
+\lambda_{2a}(N_2+1)P(N_1,N_2+1,C,Z_1,Z_2,t)
+\lambda_{1f}\sum_k^{N_1+1}\sum_l^{C}(N_1+1-k)f_{1f}(k,l)P(N_1+1-k,N_2,C-l,Z_1,Z_2,t)
+\lambda_{2f}\sum_k^{N_1}\sum_l^{C}(N_2+1)f_{2f}(k,l)P(N_1-k,N_2+1,C-l,Z_1,Z_2,t)
+\lambda_{R}(N_1+1)P(N_1+1,N_2-1,C,Z_1,Z_2,t)
+\lambda_{1d}(N_1+1)P(N_1+1,N_2,C,Z_1-1,Z_2,t)
+\lambda_{2d}(N_2+1)P(N_1,N_2+1,C,Z_1,Z_2-1,t)
+\lambda(C+1)P(N_1-1,N_2,C+1,Z_1,Z_2,t)
+S_1\sum_n^{N_1}p_q(n)P(N_1-n,N_2,C,Z_1,Z_2,t)
\end{dmath*}
with initial condition
\begin{dmath*}[style={\small}]
P(N_1,N_2,C,Z_1,Z_2,t=t_0\mid t_0) = \delta_{N_{1},0} \delta_{N_{2},0} \delta_{C,0} \delta_{Z_{1},0} \delta_{Z_{2},0}
\end{dmath*}
and
\begin{dmath*}[style={\small}]
\sum_{N_1} \sum_{N_2} \sum_{C} P(N_1,N_2,C,Z_1,Z_2,t=t_d\mid t_0) = \delta_{Z_{1},0} \delta_{Z_{2},0}
\end{dmath*}
By defining the following generating function for the probability distribution $P(N_1,N_2,C,Z_1,Z_2,t)$:
\begin{dmath*}[style={\small}]
G(X,Y,V,M,N,t)= \sum_{N_1} \sum_{N_2} \sum_{C} \sum_{Z_1}\sum_{Z_2} X^{N_1} Y^{N_2} V^{C} M^{Z_1} N^{Z_2} P(N_1,N_2,C,Z_1,Z_2,t)
\end{dmath*}
with initial condition for ${t_0 \leq 0}$
\begin{dmath*}[style={\small}]
G(X,Y,V,M,N,t=t_0\mid t_0) = 1
\end{dmath*}
and
\begin{dmath*}[style={\small}]
G(1,1,1,M,N,t=t_d\mid t_0) = 1
\end{dmath*}
the following partial differential equation is obtained:
\begin{dmath*}[style={\small}]
\frac{\partial G}{\partial t}= [\lambda_{1a}+ \lambda_{R}Y + q_1(X,V) \lambda_{1f} + \lambda_{1d} M - \lambda_1 X] \frac{\partial G}{\partial X} + [\lambda_{2a}  +  q_2(X,V)\lambda_{2f} + \lambda_{2d} N - \lambda_2 Y ] \frac{\partial G}{\partial Y} +\lambda (X-V)\frac{{\partial G}}{\partial V}+ S_1 [r(X) - 1] G,
\end{dmath*}
where
\begin{dgroup*}
\begin{dmath*}[style={\small}]
q_1 (X,V) = \sum_k \sum_l  X^k V^l f_{1f}(k,l)
\end{dmath*}
\begin{dmath*}[style={\small}]
q_2 (X,V) = \sum_k \sum_l  X^k V^l f_{2f}(k,l)
\end{dmath*}
\begin{dmath*}[style={\small}]
r(X) = \sum_n p_q(n) X^n
\end{dmath*}
\end{dgroup*}
Here, ${f_{1f}}$\emph{(k,l)} is the probability of having \emph{k} prompt neutrons and \emph{l} delay neutron precursors produced in a fission event induced by a fast neutron, ${f_{2f}}$\emph{(k,l)} is the probability of having \emph{k} prompt neutrons and \emph{l} delay neutron precursors produced in a fission event induced by a thermal neutron. The effective delayed neutron fraction is $\beta$, ${\nu_{1}'}$ and ${\nu_{2}'}$ are the average total number of neutrons per fast and thermal induced fission, respectively. For the sake of simplicity, some identities are used in the solution as below ($i=1,2$):
\begin{dgroup*}
\begin{dmath*}[style={\small}]
\left.\frac{\partial}{\partial X}q_i (X,V)\right|_{X=1,V=1}= \sum_k \sum_l  k f_{if}(k,l) =(1-\beta)\nu_{i}'
\end{dmath*}
\begin{dmath*}[style={\small}]
\left.\frac{\partial}{\partial V}q_i (X,V)\right|_{X=1,V=1}= \sum_k \sum_l  l f_{if}(k,l) =\beta\nu_{i}'
\end{dmath*}
\end{dgroup*}
and
\begin{dgroup*}
\begin{dmath*}[style={\small}]
\left.\frac{\partial}{\partial X}r (X)\right|_{X=1}= \sum_n n p_{q}(n) =r'
\end{dmath*}
\begin{dmath*}[style={\small}]
\left.\frac{\partial^2}{\partial X^2}r (X)\right|_{X=1}= \sum_n n(n-1) p_{q}(n) =r''
\end{dmath*}
\end{dgroup*}
In a steady subcritical medium with a steady source, when $t_0 \to -\infty$, the following stationary solutions for the neutron populations ${N_{1}}$, ${N_{2}}$ and ${C}$, and detection counts ${Z_{1}}$ and ${Z_{2}}$ are obtained as below:
\begin{dgroup*}
\begin{dmath*}[style={\small}]
 \bar{N}_1 = \frac{\lambda_2  S_1 r'}{\lambda_{1} \lambda_{2} - \lambda_2 \nu_1' \lambda_{1f} -  \lambda_R \nu_2' \lambda_{2f}}
\end{dmath*}
\begin{dmath*}[style={\small}]
\bar{N}_2 = \frac{\lambda_R  S_1 r'}{ \lambda_{1} \lambda_{2} - \lambda_2 \nu_1' \lambda_{1f} -  \lambda_R \nu_2' \lambda_{2f}}
\end{dmath*}
\begin{dmath*}[style={\small}]
\bar{C} = \frac{(\lambda_2 \beta\nu_1'\lambda_{1f} +\lambda _R \beta\nu_2'\lambda_{2f} ) S_1 r'}{\lambda(\lambda_{1} \lambda_{2} - \lambda_2 \nu_1' \lambda_{1f} - \lambda_R \nu_2' \lambda_{2f})}=
\frac{\bar{N}_1\beta\nu_1'\lambda_{1f}}{\lambda} +\frac{\bar{N}_2 \beta\nu_2'\lambda_{2f} }{\lambda}
\end{dmath*}
\begin{dmath*}[style={\small}]
\bar{Z}_1 =\lambda_{1d} \bar{N}_1 t
\end{dmath*}
\begin{dmath*}[style={\small}]
\bar{Z}_2 =\lambda_{2d} \bar{N}_2 t
\end{dmath*}
\end{dgroup*}
By introducing the modified second factorial moment of the random variables a and b as follows ${\mu_{aa}\equiv<a(a-1)>-<a>^{2}}$= ${\sigma_{a}^{2}}$ - ${<a>}$, ${\mu_{ab}\equiv<ab>-<a><b>}$  and then taking cross- and auto-derivatives, the following system of differential equations of modified second factorial moments for the neutron population are obtained as below:
\begin{dgroup*}
\begin{dmath*}[style={\small}]
\frac{\partial}{\partial t} \mu_{X X} =  S_1 r'' + \lambda _{2f} \nu_{2pp} \bar{N}_2 +\lambda _{1f} \nu_{1pp} \bar{N}_1 +2 \lambda  \mu_{X V}+ 2 \left[-\lambda _1+(1-\beta ) \lambda _{1f} \nu_1'\right] \mu_{X X}   +2 (1-\beta ) \lambda _{2f} \nu_2' \mu_{X Y}
\end{dmath*}
\begin{dmath*}[style={\small}]
\frac{\partial}{\partial t} \mu_{X Y} =  \lambda\mu_{Y V}+(1-\beta ) \lambda _{2f} \nu_2' \mu_{Y Y}+\left[(1-\beta)\lambda _{1f} \nu_1' -\lambda _1  -\lambda _2 \right]\mu_{X Y}+\lambda _R\mu_{X X}
\end{dmath*}
\begin{dmath*}[style={\small}]
 \frac{\partial}{\partial t} \mu_{Y Y}=-2 \lambda _2\mu_{YY} +2 \lambda _R \mu_{XY}
 \end{dmath*}
\begin{dmath*}[style={\small}]
 \frac{\partial}{\partial t} \mu_{X V}=\lambda \mu_{VV}+\lambda _{2f} \nu_{2pd} \bar{N}_2 +(1-\beta ) \lambda _{2f} \nu_2' \mu_{YV}+\lambda _{1f}\nu_{1pd} \bar{N}_1 +\left[-\lambda _1+(1-\beta ) \lambda _{1f} \nu_1'-\lambda\right] \mu_{XV}+\beta  \lambda _{2f} \nu_2' \mu_{XY}+\beta  \lambda _{1f} \nu_1' \mu_{XX}
 \end{dmath*}
\begin{dmath*}[style={\small}]
  \frac{\partial}{\partial t} \mu_{Y V}  = (-\lambda-\lambda _2 )  \mu_{YV} +\beta  \lambda _{2f} \nu_2' \mu_{YY}+\lambda _R   \mu_{XV}+\beta  \lambda _{1f} \nu_1' \mu_{XY}
\end{dmath*}
\begin{dmath*}[style={\small}]
 \frac{\partial}{\partial t} \mu_{VV}= -2 \lambda\mu_{VV}  +\lambda _{2f} \nu_{2dd}\bar{N}_2+2 \beta  \lambda _{2f} \nu_2' \mu_{YV}+\lambda _{1f} \nu_{1dd} \bar{N}_1+2 \beta  \lambda _{1f} \nu_1' \mu_{XV}
\end{dmath*}
\end{dgroup*}
The three coefficients ${\omega_1}$, ${\omega_2}$ and ${\omega_3}$ can be obtained by solving the third order equation in ${\omega}$ with known constant coefficients a, b, c:
\begin{dmath*}[style={\small}]
\omega^3+ a\cdot\omega^2+b\cdot\omega+c = 0
\end{dmath*}
where
\begin{dgroup*}
\begin{dmath*}[style={\small}]
a=  \beta  \nu _1' \lambda _{1f}-\nu _1' \lambda _{1f}+\lambda +\lambda _1+\lambda _2=-(\omega _1+ \omega _2+\omega _3)
\end{dmath*}
\begin{dmath*}[style={\small}]
b= \beta  \lambda _2 \nu _1' \lambda _{1f}-\lambda  \nu _1' \lambda _{1f}-\lambda _2 \nu _1' \lambda _{1f}+\beta  \nu _2' \lambda _{2f} \lambda _R-\nu _2' \lambda _{2f} \lambda _R+\lambda  \lambda _1+\lambda _2 \lambda _1+\lambda  \lambda _2
\end{dmath*}
\begin{dmath*}[style={\small}]
c=-\lambda  \lambda _2 \nu _1' \lambda _{1f}-\lambda  \nu _2' \lambda _{2f} \lambda _R+\lambda  \lambda _1 \lambda _2=-\omega _1 \omega _2\omega _3
\end{dmath*}
\end{dgroup*}
The stationary modified variance of the fast particle detections can be obtained from the coupled equation system by using the Laplace transform technique:
\begin{dgroup*}
\begin{dmath*}[style={\small}]
  \frac{\partial}{\partial t} \mu_{X M}  = \lambda  \mu_{VM}+(1-\beta ) \lambda _{2f} \nu_2' \mu_{YM}+\left(-\lambda _1+(1-\beta ) \lambda _{1f} \nu_1'\right) \mu_{XM}+\lambda _{1d} \mu_{XX}  \end{dmath*}
\begin{dmath*}[style={\small}]
  \frac{\partial}{\partial t} \mu_{Y M}= -\lambda _2 \mu_{YM}+\lambda _R \mu_{XM}+\lambda _{1d} \mu_{XY}
\end{dmath*}
\begin{dmath*}[style={\small}]
  \frac{\partial}{\partial t} \mu_{V M} = -\lambda  \mu_{VM}+\beta  \lambda _{2f} \nu_2' \mu_{YM}+\lambda _{1d} \mu_{XV}+\beta  \lambda _{1f} \nu_1' \mu_{XM}
\end{dmath*}
\begin{dmath*}[style={\small}]
  \frac{\partial}{\partial t} \mu_{MM} = 2 \lambda _{1d}\mu_{XM}
\end{dmath*}
\end{dgroup*}
The same can be done to define the stationary modified variance of the thermal particle detections via solving the following coupled equation system:
\begin{dgroup*}
\begin{dmath*}[style={\small}]
  \frac{\partial}{\partial t} \mu_{X N}  = \lambda  \mu_{VN}+(1-\beta ) \lambda _{2f} \nu_2' \mu_{YN}+\left(-\lambda _1+(1-\beta ) \lambda _{1f} \nu_1'\right) \mu_{XN}+\lambda _{2d} \mu_{XY}
\end{dmath*}
\begin{dmath*}[style={\small}]
  \frac{\partial}{\partial t} \mu_{Y N} = -\lambda _2 \mu_{YN}+\lambda _R \mu_{XN}+\lambda _{2d} \mu_{YY}
\end{dmath*}
\begin{dmath*}[style={\small}]
  \frac{\partial}{\partial t} \mu_{V N} = -\lambda  \mu_{VN}+\beta  \lambda _{2f} \nu_2' \mu_{YN}+\lambda _{2d} \mu_{YV}+\beta  \lambda _{1f} \nu_1' \mu_{XN}
\end{dmath*}
\begin{dmath*}[style={\small}]
 \frac{\partial}{\partial t} \mu_{NN} = 2 \lambda _{2d}\mu_{YN}
\end{dmath*}
\end{dgroup*}
Some second moment notations were introduced as follows:
\begin{dgroup*}
\begin{dmath*}[style={\small}]
\left.\frac{\partial^2}{\partial X^2}q_i (X,V)\right|_{X=1,V=1}= \sum_k \sum_l  k(k-1) f_{if}(k,l) =\nu_{ipp}
\end{dmath*}
\begin{dmath*}[style={\small}]
\left.\frac{\partial^2}{\partial V^2}q_i (X,V)\right|_{X=1,V=1}= \sum_k \sum_l  l(l-1) f_{if}(k,l) =\nu_{idd}
\end{dmath*}
\begin{dmath*}[style={\small}]
\left.\frac{\partial^2}{\partial V\partial X}q_i (X,V)\right|_{X=1,V=1}= \sum_k \sum_l  kl f_{if}(k,l) =\nu_{ipd}
\end{dmath*}
\end{dgroup*}
in which $i=1,2$.
Thus, the solution for the two-group one-point Feynman-alpha formula for fast and thermal detection particles can be written as below:
\begin{dmath*}[style={\small}]
\frac{\sigma_{ZZ}^2(t)}{\bar{Z}_1 \ \bar{Z}_2} = 1 + Y(t)=1+ \sum_{i=1}^{3} Y_i (1 - \frac{1-e^{- \omega_i t}}{\omega_i t}).
\end{dmath*}
For fast particle detections the following expressions should be used:
\begin{dgroup*}
\begin{dmath*}[style={\small}]
-Y_1=-\frac{2 \lambda _{{1d}} \left(\omega _1 \left(K_2 \omega _1-K_1\right)+K_0\right)}{\bar{N}_1 \omega _1 \left(\omega _1-\omega _2\right) \left(\omega _1-\omega _3\right)}
\end{dmath*}
\begin{dmath*}[style={\small}]
-Y_2 =\frac{2 \lambda _{{1d}} \left(\omega _2 \left(K_2 \omega _2-K_1\right)+K_0\right)}{\bar{N}_1 \left(\omega _1-\omega _2\right) \omega _2 \left(\omega _2-\omega _3\right)}
\end{dmath*}
\begin{dmath*}[style={\small}]
-Y_3 = \frac{2 \lambda _{{1d}} \left(\omega _3 \left(K_2 \omega _3-K_1\right)+K_0\right)}{\bar{N}_1 \left(\omega _1-\omega _3\right) \omega _3 \left(\omega _3-\omega _2\right)}
\end{dmath*}
\end{dgroup*}
with
\begin{dgroup*}
\begin{dmath*}[style={\small}]
K_2=\mu _{{XX}}
\end{dmath*}
\begin{dmath*}[style={\small}]
K_1=-\beta  \lambda _{{2f}} \nu _2' \mu _{{XY}}+\lambda _{{2f}} \nu _2' \mu _{{XY}}+\lambda  \mu _{{XV}}+\lambda  \mu _{{XX}}+\lambda _2 \mu _{{XX}}
\end{dmath*}
\begin{dmath*}[style={\small}]
K_0=\lambda  \lambda _{{2f}} \nu _2' \mu _{{XY}}+\lambda  \lambda _2 \mu _{{XV}}+\lambda  \lambda _2 \mu _{{XX}}
\end{dmath*}
\end{dgroup*}
It can be shown that:
\begin{dmath*}[style={\small}]
Y_0 = Y_1+Y_2 + Y_3=\frac{2 K_0 \lambda _{{1d}}}{\bar{N}_1 \omega _1 \omega _2 \omega _3}
\end{dmath*}
If a thermal neutron detector is used, then the following expressions should be considered:
\begin{dgroup*}
\begin{dmath*}[style={\small}]
-Y_1=-\frac{2 \lambda _{{2d}} \left(\omega _1 \left(L_2 \omega _1-L_1\right)+L_0\right)}{\omega _1 \left(\omega _1-\omega _2\right) \left(\omega _1-\omega _3\right) \bar{N}_2}
\end{dmath*}
\begin{dmath*}[style={\small}]
-Y_2 =\frac{2 \lambda _{{2d}} \left(\omega _2 \left(L_2 \omega _2-L_1\right)+L_0\right)}{\left(\omega _1-\omega _2\right) \omega _2 \left(\omega _2-\omega _3\right) \bar{N}_2}
\end{dmath*}
\begin{dmath*}[style={\small}]
-Y_3 = \frac{2 \lambda _{{2d}} \left(\omega _3 \left(L_2 \omega _3-L_1\right)+L_0\right)}{\left(\omega _1-\omega _3\right) \omega _3 \left(\omega _3-\omega _2\right) \bar{N}_2}
\end{dmath*}
\end{dgroup*}
with
\begin{dgroup*}
\begin{dmath*}[style={\small}]
L_2=\mu _{{YY}}
\end{dmath*}
\begin{dmath*}[style={\small}]
L_1=\beta  \lambda _{{1f}} \nu _1' \mu _{{YY}}-\lambda _{{1f}} \nu _1' \mu _{{YY}}+\lambda _R \mu _{{XY}}+\lambda  \mu _{{YY}}+\lambda _1 \mu _{{YY}}
\end{dmath*}
\begin{dmath*}[style={\small}]
L_0=-\lambda  \lambda _{{1f}} \nu _1' \mu _{{YY}}+\lambda  \lambda _R \mu _{{XY}}+\lambda  \lambda _R \mu _{{YV}}+\lambda  \lambda _1 \mu _{{YY}}
\end{dmath*}
\end{dgroup*}
It can be shown that:
\begin{dmath*}[style={\small}]
Y_0 = Y_1+Y_2 + Y_3=\frac{2 L_0 \lambda _{{2d}}}{\omega _1 \omega _2 \omega _3 \bar{N}_2}
\end{dmath*}

\subsection{One-group two-point Feynman-alpha theory (with delayed neutrons)}
    \label{sec:7}
Similarly as in the derivation of two-group one-point version of Feynman-alpha formula, in the one-group two-point Feynman-alpha theory the joint probability of having $N_A$ neutrons in region A, $N_B$ neutrons in region B, $C_A$ delayed neutron precursors presented in region A, $C_B$ delayed neutron precursors presented in region B at time $t$, $Z_A$ neutrons have been detected in region A, and $Z_B$ neutrons have been detected in region B in the system during the period of time ${t-t_{d}\geq 0}$ can be defined as $P(N_A,N_B,C_A,C_B,Z_A,Z_B,t|t_0)$.By summing up all mutually exclusive events of the particle not having or having a specific reaction within the infinitesimally small time interval d$t$, it can be written:
\begin{dmath*}[style={\small}]
\frac{\partial P(N_{A},N_{B},C_{A},C_{B},Z_{A},Z_{B},t)}{\partial t}
=-(\lambda_{A}N_{A}+\lambda_{B}N_{B}+\lambda_{Ac}C_{A}+\lambda_{Bc}C_{B}+S_{A}+S_{B})P(N_{A},N_{B},C_{A},C_{B},Z_{A},Z_{B},t)
\end{dmath*}
\begin{dmath*}[style={\small}]
+\lambda_{Aa}(N_{A}+1)P(N_{A}+1,N_{B},C_{A},C_{B},Z_{A},Z_{B},t)
+\lambda_{Ba}(N_{B}+1)P(N_{A},N_{B}+1,C_{A},C_{B},Z_{A},Z_{B},t)
+\lambda_{Af}\sum_k^{N_{A}+1}\sum_l^{C_{A}}(N_{A}+1-k)f_{A}(k,l)P(N_{A}+1-k,N_{B},C_{A}-l,C_{B},Z_{A},Z_{B},t)
+\lambda_{Bf}\sum_k^{N_{B}+1}\sum_l^{C_{B}}(N_{B}+1-k)f_{B}(k,l)P(N_{A},N_{B}+1-k,C_{A},C_{B}-l,Z_{A},Z_{B},t)
+\lambda_{At}(N_{A}+1)P(N_{A}+1,N_{B}-1,C_{A},C_{B},Z_{A},Z_{B},t)
+\lambda_{Bt}(N_{B}+1)P(N_{A}-1,N_{B}+1,C_{A},C_{B},Z_{A},Z_{B},t)
+\lambda_{Ac}(C_{A}+1)P(N_{A}-1,N_{B},C_{A}+1,C_{B},Z_{A},Z_{B},t)
+\lambda_{Bc}(C_{B}+1)P(N_{A},N_{B}-1,C_{A},C_{B}+1,Z_{A},Z_{B},t)
+\lambda_{Ad}(N_{A}+1)P(N_{A}+1,N_{B},C_{A},C_{B},Z_{A}-1,Z_{B},t)
+\lambda_{Bd}(N_{B}+1)P(N_{A},N_{B}+1,C_{A},C_{B},Z_{A},Z_{B}-1,t)
+S_{A}\sum_n^{N_{A}}p_{A}(n)P(N_{A}-n,N_{B},C_{A},C_{B},Z_{A},Z_{B},t)
+S_{B}\sum_n^{N_{B}}p_{B}(n)P(N_{A},N_{B}-n,C_{A},C_{B},Z_{A},Z_{B},t)
\end{dmath*}
with initial condition
\begin{dmath*}[style={\small}]
P(N_{A},N_{B},C_{A},C_{B},Z_{A},Z_{B},t=t_0\mid t_0) = \delta_{N_{A},0} \delta_{N_{B},0} \delta_{C_{A},0} \delta_{C_{B},0} \delta_{Z_{A},0} \delta_{Z_{B},0}
\end{dmath*}
and
\begin{dmath*}[style={\small}]
\sum_{N_A} \sum_{N_B} \sum_{C_A}\sum_{C_B} P(N_{A},N_{B},C_{A},C_{B},Z_{A},Z_{B},t=t_d\mid t_0) = \delta_{Z_{A},0} \delta_{Z_{B},0}
\end{dmath*}
By defining the following generating function for the probability distribution $P(N_{A},N_{B},C_{A},C_{B},Z_{A},Z_{B},t)$:
\begin{dmath*}[style={\small}]
G(X,Y,V,W,M,N,t) = \sum_{N_A} \sum_{N_B} \sum_{C_A}\sum_{C_B} \sum_{Z_A}\sum_{Z_B} X^{N_A} Y^{N_B} V^{C_A}W^{C_B} M^{Z_A} N^{Z_B}P(N_A,N_B,C_A,C_B,Z_A,Z_B,t)
\end{dmath*}
with initial condition for ${t_0 \leq 0}$
\begin{dmath*}[style={\small}]
G(X,Y,V,W,M,N,t=t_0\mid t_0) = 1
\end{dmath*}
and
\begin{dmath*}[style={\small}]
G(1,1,1,1,M,N,t=t_d\mid t_0) = 1
\end{dmath*}
a partial differential equation in the variables $(X,Y,V,W,M,N)$ in terms of generating function can be obtained:
\begin{dmath*}[style={\small}]
\frac{\partial G}{\partial t}=[\lambda_{Aa}+\lambda_{At}Y + q_A(X,V) \lambda_{Af} + \lambda_{Ad} M - \lambda_A X] \frac{\partial G}{\partial X}
+ [\lambda_{Ba} +\lambda_{Bt}X  +  q_B(Y,W)\lambda_{Bf} + \lambda_{Bd} N - \lambda_B Y ] \frac{\partial G}{\partial Y}
+\lambda_{Ac} (X-V)\frac{{\partial G}}{\partial V} + \lambda_{Bc} (Y-W)\frac{{\partial G}}{\partial W}
+ S_A [r_A(X) - 1] G +S_B [r_B(Y) - 1]G
\end{dmath*}
where
\begin{dgroup*}
\begin{dmath*}[style={\small}]
q_A (X,V) = \sum_k \sum_l  X^k V^l f_{A}(k,l)
\end{dmath*}
\begin{dmath*}[style={\small}]
q_B (Y,W) = \sum_k \sum_l  Y^k W^l f_{B}(k,l)
\end{dmath*}
\begin{dmath*}[style={\small}]
r_A(X)=\sum_n p_A(n) X^n
\end{dmath*}
\begin{dmath*}[style={\small}]
r_B(Y) = \sum_n p_B(n) Y^n
\end{dmath*}
\end{dgroup*}
Here, $\beta_A$ and $\beta_B$ are the effective delayed neutron fractions in region A and region B, respectively. For the sake of simplicity, some identities are used in the solution as below ($i=A,B$):
\begin{dgroup*}
\begin{dmath*}[style={\small}]
\left.\frac{\partial}{\partial X}q_A (X,V)\right|_{X=1,V=1}= \sum_k \sum_l  k f_{Af}(k,l) =(1-\beta_A)\nu_{A}'
\end{dmath*}
\begin{dmath*}[style={\small}]
\left.\frac{\partial}{\partial V}q_A (X,V)\right|_{X=1,V=1}= \sum_k \sum_l  l f_{Af}(k,l) =\beta_A\nu_{A}'
\end{dmath*}
\begin{dmath*}[style={\small}]
\left.\frac{\partial}{\partial Y}q_B (Y,W)\right|_{Y=1,W=1}= \sum_k \sum_l  k f_{Bf}(k,l) =(1-\beta_B)\nu_{B}'
\end{dmath*}
\begin{dmath*}[style={\small}]
\left.\frac{\partial}{\partial W}q_B (Y,W)\right|_{Y=1,W=1}= \sum_k \sum_l  l f_{Bf}(k,l) =\beta_B\nu_{B}'
\end{dmath*}
\end{dgroup*}
and
\begin{dgroup*}
\begin{dmath*}[style={\small}]
\left.\frac{\partial}{\partial X}r_A (X)\right|_{X=1}= \sum_n n p_{A}(n) =r_{A}'
\end{dmath*}
\begin{dmath*}[style={\small}]
\left.\frac{\partial^2}{\partial X^2}r_A (X)\right|_{X=1}= \sum_n n(n-1) p_{A}(n) =r_{A}''
\end{dmath*}
\begin{dmath*}[style={\small}]
\left.\frac{\partial}{\partial Y}r_B (Y)\right|_{Y=1}= \sum_n n p_{B}(n) =r_{B}'
\end{dmath*}
\begin{dmath*}[style={\small}]
\left.\frac{\partial^2}{\partial Y^2}r_B (Y)\right|_{Y=1}=\sum_n n(n-1) p_{B}(n) =r_{B}''
\end{dmath*}
\end{dgroup*}
In a steady subcritical medium with a steady source, when $t_0 \to -\infty$, the following stationary solutions for the neutron population and detection counts are obtained as follows:
\begin{dgroup*}
\begin{dmath*}[style={\small}]
\bar{N}_A =   \frac{S_A \left(\lambda _B-\lambda _{ {Bf}} \nu_{B}'\right) r_A'+S_B \lambda _{ {Bt}} r_B'}{\left(\lambda_A-\lambda_{ {Af}}  \nu_A'\right)\left(\lambda_B-\lambda_{ {Bf}} \nu_B'\right)-\lambda _{ {At}} \lambda _{ {Bt}}}
\end{dmath*}
\begin{dmath*}[style={\small}]
\bar{N}_B   =\frac{S_A \lambda _{ {At}} r_A'+S_B \left(\lambda _A-\lambda _{ {Af}} \nu_{A}'\right) r_B'}{\left(\lambda_A-\lambda_{ {Af}}  \nu_A'\right)\left(\lambda_B-\lambda_{ {Bf}} \nu_B'\right)-\lambda _{ {At}} \lambda _{ {Bt}}}
\end{dmath*}
\begin{dmath*}[style={\small}]
\bar{C}_A =  \frac{\beta _A \lambda _{ {Af}} \nu_{A}'}{\lambda _{ {Ac}}}\bar{N}_A
\end{dmath*}
\begin{dmath*}[style={\small}]
\bar{C}_B =  \frac{\beta _B \lambda _{ {Bf}} \nu_{B}'}{\lambda _{ {Bc}}} \bar{N}_B
\end{dmath*}
\begin{dmath*}[style={\small}]
\bar{Z}_A = \lambda_{1d} \bar{N}_1 t
\end{dmath*}
\begin{dmath*}[style={\small}]
\bar{Z}_B = \lambda_{2d} \bar{N}_2 t
\end{dmath*}
\end{dgroup*}
By introducing the modified second factorial moments and then taking cross- and auto-derivatives, the following system of differential equations of modified second factorial moments for the neutron population are obtained as below:
\begin{dgroup*}
\begin{dmath*}[style={\small}]
\frac{\partial}{\partial t} \mu_{X X} = 2 \left[(1-\beta_A ) \lambda _{Af} \nu_{A}'-\lambda _A\right] \mu_{XX} + 2 \lambda_{Bt} \mu_{XY} +  2 \lambda_{Ac}\mu_{XV}+  S_A r_A''+   \lambda _{Af} \nu_{App}\bar{N}_A
\end{dmath*}
\begin{dmath*}[style={\small}]
  \frac{\partial}{\partial t} \mu_{X Y} =\lambda _{ {Ac}}\mu_{YV}+\lambda _{ {Bc}}\mu_{XW}+\lambda _{ {Bt}}\mu_{YY}+\lambda _{ {At}}\mu_{XX}+\left[\left(1-\beta _B\right) \lambda _{ {Bf}} \nu_{B}' -\lambda _B   +\left(1-\beta _A\right) \lambda _{ {Af}} \nu_{A}'-\lambda _A \right] \mu_{XY}
\end{dmath*}
\begin{dmath*}[style={\small}]
\frac{\partial}{\partial t} \mu_{Y Y} = 2 \left[-\lambda _B+(1-\beta_B ) \lambda _{Bf} \nu_{B}'\right] \mu_{YY} + 2 \lambda_{At} \mu_{XY} +  2 \lambda_{Bc}\mu_{YW}+  S_B r_B''+   \lambda _{Bf} \nu_{Bpp} \bar{N}_B
\end{dmath*}
\begin{dmath*}[style={\small}]
 \frac{\partial}{\partial t} \mu_{XV}= ((1-\beta_A) \nu_{A}' \lambda_{{Af}} - \lambda_A\ - \lambda_{{Ac}}) \mu_{XV}+\beta _A \nu_{A}' \lambda _{ {Af}} \mu_{XX}+\lambda _{ {Ac}} \mu_{VV} +\lambda _{ {Bt}} \mu_{YV}+\lambda _{ {Af}} \nu_{Apd} \bar{N}_A
\end{dmath*}
\begin{dmath*}[style={\small}]
  \frac{\partial}{\partial t} \mu_{YV}  = \lambda _{ {Bc}} \mu_{VW}+\left[\left(1-\beta _B\right) \lambda _{ {Bf}} \nu_{B}' - \lambda_B\ - \lambda_{{Ac}} \right]\mu_{YV} +\lambda _{ {At}} \mu_{XV}+\beta _A \lambda _{ {Af}} \nu_{A}' \mu_{XY}
\end{dmath*}
\begin{dmath*}[style={\small}]
 \frac{\partial}{\partial t} \mu_{VV}  =-2 \lambda _{ {Ac}}\mu_{VV}+2 \beta _A \lambda _{ {Af}} \nu_{A}' \mu_{XV} +\lambda _{ {Af}} \nu_{Add} \bar{N}_A
\end{dmath*}
\begin{dmath*}[style={\small}]
 \frac{\partial}{\partial t} \mu_{XW}=\lambda _{ {Ac}} \mu_{VW}+\lambda _{ {Bt}} \mu_{YW}+\left[\left(1-\beta _A\right) \lambda _{ {Af}} \nu_{A}'-\lambda _A -\lambda _{ {Bc}} \right] \mu_{XW} +\beta _B \lambda _{ {Bf}} \nu_{B}' \mu_{XY}
\end{dmath*}
\begin{dmath*}[style={\small}]
  \frac{\partial}{\partial t} \mu_{YW}= \lambda _{ {Bc}}\mu_{WW}   +\left[\left(1-\beta _B\right) \lambda _{ {Bf}} \nu_{B}' -\lambda _B  -\lambda _{ {Bc}} \right]\mu_{YW} +\beta _B \lambda _{ {Bf}} \nu_{B}' \mu_{YY}  +\lambda _{ {At}}\mu_{XW}  +\lambda _{ {Bf}} \nu_{Bpd} \bar{N}_B
\end{dmath*}
\begin{dmath*}[style={\small}]
   \frac{\partial}{\partial t} \mu_{VW}= \left(-\lambda _{ {Ac}}-\lambda _{ {Bc}} \right) \mu_{VW}  +\beta _B \lambda _{ {Bf}} \nu_{B}' \mu_{YV}+\beta _A \lambda _{ {Af}} \nu_{A}' \mu_{XW}   \end{dmath*}
\begin{dmath*}[style={\small}]
    \frac{\partial}{\partial t} \mu_{WW}= -2 \lambda _{ {Bc}} \mu_{WW}   +2 \beta _B \lambda _{ {Bf}} \nu_{B}' \mu_{YW}  +\lambda _{ {Bf}} \nu_{Bdd} \bar{N}_B
\end{dmath*}
\end{dgroup*}
where
\begin{dgroup*}
\begin{dmath*}[style={\small}]
\left.\frac{\partial^2}{\partial X^2}q_A (X,V)\right|_{X=1,V=1}= \sum_k \sum_l  k(k-1) f_{Af}(k,l) =\nu_{App}
\end{dmath*}
\begin{dmath*}[style={\small}]
\left.\frac{\partial^2}{\partial Y^2}q_B (Y,W)\right|_{Y=1,W=1}= \sum_k \sum_m  k(k-1) f_{Bf}(k,m) =\nu_{Bpp}
\end{dmath*}
\begin{dmath*}[style={\small}]
\left.\frac{\partial^2}{\partial V^2}q_A (X,V)\right|_{X=1,V=1}= \sum_k \sum_l  l(l-1) f_{Af}(k,l) =\nu_{Add}
\end{dmath*}
\begin{dmath*}[style={\small}]
\left.\frac{\partial^2}{\partial V^2}q_B (Y,W)\right|_{Y=1,W=1}= \sum_k \sum_l l(l-1) f_{Bf}(k,l) =\nu_{Bdd}
\end{dmath*}
\begin{dmath*}[style={\small}]
\left.\frac{\partial^2}{\partial V\partial X}q_A (X,V)\right|_{X=1,V=1}= \sum_k \sum_l  kl f_{Af}(k,l) =\nu_{Apd}
\end{dmath*}
\begin{dmath*}[style={\small}]
\left.\frac{\partial^2}{\partial W\partial Y}q_B (Y,W)\right|_{Y=1,W=1}= \sum_k \sum_l  kl f_{Bf}(k,l) =\nu_{Bpd}
\end{dmath*}
\end{dgroup*}
The system above is solved for stationary case when ${\frac{\partial}{\partial t}=0}$.
Four roots ${\omega_1}$, ${\omega_2}$, ${\omega_3}$ and ${\omega_4}$ can be obtained by solving the forth order equation with coefficients a, b, c, d specified as below:
\begin{dgroup*}
\begin{dmath*}[style={\small}]
\omega^4+ a\cdot\omega^3+b\cdot\omega^2+c\cdot\omega+d = 0
\end{dmath*}
\begin{dmath*}[style={\small}]
a=\beta _A \lambda _{{Af}} \nu _A'-\lambda _{{Af}} \nu _A'+\lambda _A+\lambda _{{Ac}}+\beta _B \lambda _{{Bf}} \nu _B'-\lambda _{{Bf}} \nu _B'+\lambda _B+\lambda _{{Bc}}
\end{dmath*}
\begin{dmath*}[style={\small}]
b=-\lambda _{{Ac}} \lambda _{{Af}} \nu _A'+\lambda _A \lambda _{{Ac}}+\beta _A \lambda _{{Af}} \lambda _B \nu _A'-\beta _A \lambda _{{Af}} \lambda _{{Bf}} \nu _A' \nu _B'+\beta _A \lambda _{{Af}} \beta _B \lambda _{{Bf}} \nu _A' \nu _B'-\lambda _{{Af}} \beta _B \lambda _{{Bf}} \nu _A' \nu _B'+\lambda _{{Af}} \lambda _{{Bf}} \nu _A' \nu _B'-\lambda _{{Af}} \lambda _B \nu _A'+\beta _A \lambda _{{Af}} \lambda _{{Bc}} \nu _A'-\lambda _{{Af}} \lambda _{{Bc}} \nu _A'+\lambda _A \beta _B \lambda _{{Bf}} \nu _B'-\lambda _A \lambda _{{Bf}} \nu _B'+\lambda _A \lambda _B+\lambda _A \lambda _{{Bc}}+\lambda _{{Ac}} \beta _B \lambda _{{Bf}} \nu _B'-\lambda _{{Ac}} \lambda _{{Bf}} \nu _B'+\lambda _{{Ac}} \lambda _B+\lambda _{{Ac}} \lambda _{{Bc}}-\lambda _{{At}} \lambda _{{Bt}}-\lambda _{{Bc}} \lambda _{{Bf}} \nu _B'+\lambda _B \lambda _{{Bc}}
\end{dmath*}
\begin{dmath*}[style={\small}]
c=-\lambda _{{Ac}} \lambda _{{Af}} \beta _B \lambda _{{Bf}} \nu _A' \nu _B'+\lambda _{{Ac}} \lambda _{{Af}} \lambda _{{Bf}} \nu _A' \nu _B'-\lambda _{{Ac}} \lambda _{{Af}} \lambda _B \nu _A'-\lambda _{{Ac}} \lambda _{{Af}} \lambda _{{Bc}} \nu _A'+\lambda _A \lambda _{{Ac}} \beta _B \lambda _{{Bf}} \nu _B'-\lambda _A \lambda _{{Ac}} \lambda _{{Bf}} \nu _B'+\lambda _A \lambda _{{Ac}} \lambda _B+\lambda _A \lambda _{{Ac}} \lambda _{{Bc}}+\beta _A \lambda _{{Af}} \lambda _B \lambda _{{Bc}} \nu _A'-\beta _A \lambda _{{Af}} \lambda _{{Bc}} \lambda _{{Bf}} \nu _A' \nu _B'+\lambda _{{Af}} \lambda _{{Bc}} \lambda _{{Bf}} \nu _A' \nu _B'-\lambda _{{Af}} \lambda _B \lambda _{{Bc}} \nu _A'-\lambda _A \lambda _{{Bc}} \lambda _{{Bf}} \nu _B'+\lambda _A \lambda _B \lambda _{{Bc}}-\lambda _{{Ac}} \lambda _{{At}} \lambda _{{Bt}}-\lambda _{{Ac}} \lambda _{{Bc}} \lambda _{{Bf}} \nu _B'+\lambda _{{Ac}} \lambda _B \lambda _{{Bc}}-\lambda _{{At}} \lambda _{{Bc}} \lambda _{{Bt}}
\end{dmath*}
\begin{dmath*}[style={\small}]
d=\lambda _{{Ac}} \lambda _{{Af}} \lambda _{{Bc}} \lambda _{{Bf}} \nu _A' \nu _B'-\lambda _{{Ac}} \lambda _{{Af}} \lambda _B \lambda _{{Bc}} \nu _A'-\lambda _A \lambda _{{Ac}} \lambda _{{Bc}} \lambda _{{Bf}} \nu _B'+\lambda _A \lambda _{{Ac}} \lambda _B \lambda _{{Bc}}-\lambda _{{Ac}} \lambda _{{At}} \lambda _{{Bc}} \lambda _{{Bt}}
\end{dmath*}
\end{dgroup*}
The stationary modified variance of the particle detections in Region A can be obtained from the coupled equation system by using the Laplace transform technique:
\begin{dgroup*}
\begin{dmath*}[style={\small}]
  \frac{\partial}{\partial t} \mu_{XM}  =\lambda _{ {Ac}}\mu_{VM} +\lambda _{ {Bt}} \mu_{YM}  +\left[\left(1-\beta _A\right) \lambda _{ {Af}} \nu_{A}'-\lambda _A\right]\mu_{XM}     +\lambda _{ {Ad}} \mu_{XX}
\end{dmath*}
\begin{dmath*}[style={\small}]
  \frac{\partial}{\partial t} \mu_{YM}=   \lambda _{ {Bc}} \mu_{WM}+\left[\left(1-\beta _B\right) \lambda _{ {Bf}} \nu_{B}'-\lambda _B\right] \mu_{YM}+\lambda _{ {At}} \mu_{XM}+\lambda _{ {Ad}} \mu_{XY}
\end{dmath*}
\begin{dmath*}[style={\small}]
  \frac{\partial}{\partial t} \mu_{VM} =-\lambda _{ {Ac}}\mu_{VM}  +\beta _A \lambda _{ {Af}} \nu_{A}' \mu_{XM} +\lambda _{ {Ad}} \mu_{XV}
\end{dmath*}
\begin{dmath*}[style={\small}]
    \frac{\partial}{\partial t} \mu_{WM} = -\lambda _{ {Bc}} \mu_{WM}+\beta _B \lambda _{ {Bf}} \nu_{B}' \mu_{YM}+\lambda _{ {Ad}} \mu_{XW}
\end{dmath*}
\begin{dmath*}[style={\small}]
  \frac{\partial}{\partial t} \mu_{MM} = 2 \lambda _{Ad}\mu_{XM}
\end{dmath*}
\end{dgroup*}
A similar coupled equation system can be derived for the particle detections in Region B:
\begin{dgroup*}
\begin{dmath*}[style={\small}]
  \frac{\partial}{\partial t} \mu_{XN}=\lambda _{ {Ac}} \mu_{VN}+\lambda _{ {Bt}} \mu_{YN}+\left[\left(1-\beta _A\right) \lambda _{ {Af}} \nu_{A}'-\lambda _A\right] \mu_{XN}+\lambda _{ {Bd}}\mu_{XY}
  \end{dmath*}
\begin{dmath*}[style={\small}]
  \frac{\partial}{\partial t} \mu_{YN}=\lambda _{ {Bc}}\mu_{WN}   +\left[\left(1-\beta _B\right) \lambda _{ {Bf}} \nu_{B}'-\lambda _B\right] \mu_{YN}+\lambda _{ {At}} \mu_{XN} +\lambda _{ {Bd}} \mu_{YY}
\end{dmath*}
\begin{dmath*}[style={\small}]
  \frac{\partial}{\partial t} \mu_{VN}= -\lambda _{ {Ac}}\mu_{VN} +\beta _A \lambda _{ {Af}} \nu_{A}' \mu_{XN}   +\lambda _{ {Bd}} \mu_{YV}
\end{dmath*}
\begin{dmath*}[style={\small}]
    \frac{\partial}{\partial t} \mu_{WN}= -\lambda _{ {Bc}} \mu_{WN}   +\beta _B \lambda _{ {Bf}} \nu_{B}' \mu_{YN}   +\lambda _{ {Bd}} \mu_{YW}
\end{dmath*}
\begin{dmath*}[style={\small}]
 \frac{\partial}{\partial t} \mu_{NN} = 2 \lambda _{Bd}\mu_{YN}
\end{dmath*}
\end{dgroup*}
Thus, a final expression for the two-point one-group Feynman-alpha formula for region A and B is written below:
\begin{dmath*}[style={\small}]
\frac{\sigma_{ZZ}^2(t)}{\bar{Z}_A/\bar{Z}_B} =1+Y(t)+ 1 +\sum_{i=1}^{4} Y_i (1 - \frac{1-e^{- \omega_i t}}{\omega_i t})
\end{dmath*}
If the detector is placed in Region A, the following expressions for the functions ${Y_i}$ should be used:
\begin{dgroup*}
\begin{dmath*}[style={\small}]
  -Y_1 = \frac{2 \lambda _{{Ad}} \left(K_0-\omega _1 \left(\omega _1 \left(K_3 \omega _1-K_2\right)+K_1\right)\right)}{\bar{N}_A \omega _1 \left(\omega _1-\omega _2\right) \left(\omega _1-\omega _3\right) \left(\omega _1-\omega _4\right)}
\end{dmath*}
\begin{dmath*}[style={\small}]
    -Y_2 = \frac{2 \lambda _{{Ad}} \left(K_0-\omega _2 \left(\omega _2 \left(K_3 \omega _2-K_2\right)+K_1\right)\right)}{\bar{N}_A \omega _2 \left(\omega _2-\omega _1\right) \left(\omega _2-\omega _3\right) \left(\omega _2-\omega _4\right)}
\end{dmath*}
\begin{dmath*}[style={\small}]
    - Y_3 = \frac{2 \lambda _{{Ad}} \left(K_0-\omega _3 \left(\omega _3 \left(K_3 \omega _3-K_2\right)+K_1\right)\right)}{\bar{N}_A \omega _3 \left(\omega _3-\omega _1\right) \left(\omega _3-\omega _2\right) \left(\omega _3-\omega _4\right)}
\end{dmath*}
\begin{dmath*}[style={\small}]
     - Y_4 = \frac{2 \lambda _{{Ad}} \left(K_0-\omega _4 \left(\omega _4 \left(K_3 \omega _4-K_2\right)+K_1\right)\right)}{\bar{N}_A \omega _4 \left(\omega _4-\omega _1\right) \left(\omega _4-\omega _2\right) \left(\omega _4-\omega _3\right)}
\end{dmath*}
\end{dgroup*}
and it can be proved that:
\begin{dmath*}[style={\small}]
  Y_0 =Y_1+Y_2 + Y_3+ Y_4=\frac{2 K_0 \lambda _{{Ad}}}{ \bar{N}_A \omega _1 \omega _2 \omega _3 \omega _4}
\end{dmath*}
where
\begin{dgroup*}
\begin{dmath*}[style={\small}]
  K_3=\mu _{{XX}}
\end{dmath*}
\begin{dmath*}[style={\small}]
K_2=\lambda _{{Ac}} \mu _{{XV}}+\lambda _{{Ac}} \mu _{{XX}}+\beta _B \lambda _{{Bf}} \mu _{{XX}} \nu _B'-\lambda _{{Bf}} \mu _{{XX}} \nu _B'+\lambda _B \mu _{{XX}}+\lambda _{{Bc}} \mu _{{XX}}+\lambda _{{Bt}} \mu _{{XY}}
\end{dmath*}
\begin{dmath*}[style={\small}]
 K_1=\lambda _{{Ac}} \beta _B \lambda _{{Bf}} \mu _{{XV}} \nu _B'-\lambda _{{Ac}} \lambda _{{Bf}} \mu _{{XV}} \nu _B'+\lambda _{{Ac}} \beta _B \lambda _{{Bf}} \mu _{{XX}} \nu _B'-\lambda _{{Ac}} \lambda _{{Bf}} \mu _{{XX}} \nu _B'+\lambda _{{Ac}} \lambda _B \mu _{{XV}}+\lambda _{{Ac}} \lambda _B \mu _{{XX}}+\lambda _{{Ac}} \lambda _{{Bc}} \mu _{{XV}}+\lambda _{{Ac}} \lambda _{{Bc}} \mu _{{XX}}+\lambda _{{Ac}} \lambda _{{Bt}} \mu _{{XY}}-\lambda _{{Bc}} \lambda _{{Bf}} \mu _{{XX}} \nu _B'+\lambda _B \lambda _{{Bc}} \mu _{{XX}}+\lambda _{{Bc}} \lambda _{{Bt}} \mu _{{XW}}+\lambda _{{Bc}} \lambda _{{Bt}} \mu _{{XY}}
\end{dmath*}
\begin{dmath*}[style={\small}]
K_0=-\lambda _{{Ac}} \lambda _{{Bc}} \lambda _{{Bf}} \mu _{{XV}} \nu _B'-\lambda _{{Ac}} \lambda _{{Bc}} \lambda _{{Bf}} \mu _{{XX}} \nu _B'+\lambda _{{Ac}} \lambda _B \lambda _{{Bc}} \mu _{{XV}}+\lambda _{{Ac}} \lambda _B \lambda _{{Bc}} \mu _{{XX}}+\lambda _{{Ac}} \lambda _{{Bc}} \lambda _{{Bt}} \mu _{{XW}}+\lambda _{{Ac}} \lambda _{{Bc}} \lambda _{{Bt}} \mu _{{XY}}
\end{dmath*}
\end{dgroup*}
If the detector is placed in Region B, the following expressions for the functions ${Y_i}$ should be used:
\begin{dgroup*}
\begin{dmath*}[style={\small}]
-Y_1 = \frac{2 \lambda _{{Bd}} \left(L_0-\omega _1 \left(\omega _1 \left(L_3 \omega _1-L_2\right)+L_1\right)\right)}{\omega _1 \left(\omega _1-\omega _2\right) \left(\omega _1-\omega _3\right) \left(\omega _1-\omega _4\right) \bar{N}_B}
\end{dmath*}
\begin{dmath*}[style={\small}]
-Y_2 = \frac{2 \lambda _{{Bd}} \left(L_0-\omega _2 \left(\omega _2 \left(L_3 \omega _2-L_2\right)+L_1\right)\right)}{\omega _2 \left(\omega _2-\omega _1\right) \left(\omega _2-\omega _3\right) \left(\omega _2-\omega _4\right) \bar{N}_B}
\end{dmath*}
\begin{dmath*}[style={\small}]
- Y_3 =\frac{2 \lambda _{{Bd}} \left(L_0-\omega _3 \left(\omega _3 \left(L_3 \omega _3-L_2\right)+L_1\right)\right)}{\omega _3 \left(\omega _3-\omega _1\right) \left(\omega _3-\omega _2\right) \left(\omega _3-\omega _4\right)\bar{N}_B}
\end{dmath*}
\begin{dmath*}[style={\small}]
- Y_4 = \frac{2 \lambda _{{Bd}} \left(L_0-\omega _4 \left(\omega _4 \left(L_3 \omega _4-L_2\right)+L_1\right)\right)}{\omega _4 \left(\omega _4-\omega _1\right) \left(\omega _4-\omega _2\right) \left(\omega _4-\omega _3\right) \bar{N}_B}
\end{dmath*}
\end{dgroup*}
and it can be proved that:
\begin{dmath*}[style={\small}]
  Y_0 =Y_1+Y_2 + Y_3+ Y_4=\frac{2 L_0 \lambda _{{Bd}}}{\omega _1 \omega _2 \omega _3 \omega _4 \bar{N}_B}
\end{dmath*}
where
\begin{dgroup*}
\begin{dmath*}[style={\small}]
L_3=\mu _{{YY}}
\end{dmath*}
\begin{dmath*}[style={\small}]
L_2=\beta _A \lambda _{{Af}} \mu _{{YY}} \nu _A'-\lambda _{{Af}} \mu _{{YY}} \nu _A'+\lambda _A \mu _{{YY}}+\lambda _{{Ac}} \mu _{{YY}}+\lambda _{{At}} \mu _{{XY}}+\lambda _{{Bc}} \mu _{{YW}}+\lambda _{{Bc}} \mu _{{YY}}
\end{dmath*}
\begin{dmath*}[style={\small}]
L_1=-\lambda _{{Ac}} \lambda _{{Af}} \mu _{{YY}} \nu _A'+\lambda _A \lambda _{{Ac}} \mu _{{YY}}+\beta _A \lambda _{{Af}} \lambda _{{Bc}} \mu _{{YW}} \nu _A'-\lambda _{{Af}} \lambda _{{Bc}} \mu _{{YW}} \nu _A'+\beta _A \lambda _{{Af}} \lambda _{{Bc}} \mu _{{YY}} \nu _A'-\lambda _{{Af}} \lambda _{{Bc}} \mu _{{YY}} \nu _A'+\lambda _A \lambda _{{Bc}} \mu _{{YW}}+\lambda _A \lambda _{{Bc}} \mu _{{YY}}+\lambda _{{Ac}} \lambda _{{At}} \mu _{{XY}}+\lambda _{{Ac}} \lambda _{{At}} \mu _{{YV}}+\lambda _{{Ac}} \lambda _{{Bc}} \mu _{{YW}}+\lambda _{{Ac}} \lambda _{{Bc}} \mu _{{YY}}+\lambda _{{At}} \lambda _{{Bc}} \mu _{{XY}}
\end{dmath*}
\begin{dmath*}[style={\small}]
L_0=-\lambda _{{Ac}} \lambda _{{Af}} \lambda _{{Bc}} \mu _{{YW}} \nu _A'-\lambda _{{Ac}} \lambda _{{Af}} \lambda _{{Bc}} \mu _{{YY}} \nu _A'+\lambda _A \lambda _{{Ac}} \lambda _{{Bc}} \mu _{{YW}}+\lambda _A \lambda _{{Ac}} \lambda _{{Bc}} \mu _{{YY}}+\lambda _{{Ac}} \lambda _{{At}} \lambda _{{Bc}} \mu _{{XY}}+\lambda _{{Ac}} \lambda _{{At}} \lambda _{{Bc}} \mu _{{YV}}
\end{dmath*}
\end{dgroup*}

\section{Discussion and quantitative analysis}
In the following, we shall perform a comparison of the two-point two-group version of the Feynman-alpha theoretical formula to the two-point one-group, the one-point two-group and the one-point one-group (i.e. traditional) versions.
     \label{sec:9}
      \subsection{The simulation set-up}
      \label{sec:10}
\begin{figure}[ht!]
\centering
\includegraphics[width=0.35\textwidth]{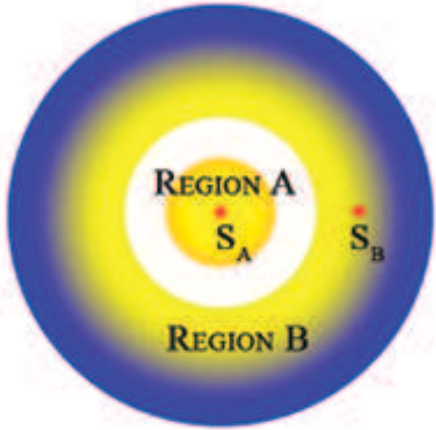}
\caption{Geometry used for the Monte-Carlo simulations.}
\label{fig:2}
\end{figure}

In order to compare the four different versions of the Feynman-alpha theory, quantitative values of the transition probabilities and reaction intensities were obtained by using Monte-Carlo simulations in a way similar to that described in  \cite{Chernikova2013,Anderson20121,Anderson2012,Chernikova20131}. The simulation setup consists of two regions, Region A and Region B, as shown in Figure \ref{fig:2}.
Region A represents nuclear material (radius 4.46 cm), in particular a mixture of 2.5\% ${^{235}}$U and 97.5\% ${^{238}}$U. Region B consists of a moderating material with a thickness of 21 cm. The neutron source emits the neutrons with an energy of 2.5 MeV. Two cases are considered in the simulations, one when the neutron source is in the center of Region A and another when the neutron source is at a distance of 15 cm from the center of the nuclear material, in Region B. Two point detectors, in Region A and in Region B, are included in the simulation setup. Delay neutron precursors are not included in the simulations.
\subsection{Coefficients}
      \label{sec:11}
Initially, the transition probabilities and reaction intensities are obtained in simulations for the two-point two-group case. Then, the values of reaction intensities of the two-point two-group case are condensed in order to get the reaction intensities which correspond to the two-point one-group, the one-point two-group and one-point one-group cases. Afterwards, these values are used in order to obtain the values of the $Y$ and $\omega$ coefficients in the Feynman-alpha formulas for the cases when the source is in Region A, and in Region B for fast neutron detections and thermal neutron detections, as shown in Table \ref{table:tab1}-\ref{table:tab4}, respectively.
\begin{table}[htbp]\footnotesize
\caption{The values of the $Y_i$ and $\omega_i$ calculated for four various versions of Feynman-alpha formulas (the source is in Region A, fast neutron detector is used either in Region A or B).}
  \centering
    \begin{tabular}{rrrrr}
        \textbf{} & \multicolumn{4}{c}{\textbf{Source in Region A}} \\
          & \textbf{2-point} & \textbf{2-point } & \textbf{1-point} & \textbf{1-point} \\
    \textbf{} & \textbf{2-group} & \textbf{1-group} & \textbf{2-group} & \textbf{1-group} \\
    $\omega_1$    & 1.52001 &1.8083 & 1.45471 & 0.923611 \\
    $\omega_2$     & 1.12141 &0.743335 & 0.350087 &  \\
    $\omega_3$    & 0.759289 &       &       &  \\
    $\omega_4$     & 0.0983484&       &       &  \\
    \textbf{} & \multicolumn{4}{c}{\textbf{Fast neutron detections in Region A}} \\
    $Y_1$    & 0.00239792 & 0.00154341 & 0.000823734 & 0.0126651 \\
    $Y_2$    & 0.00354816 & 0.00938149 & 0.0544565 &  \\
    $Y_3$    & 0.00061659 &       &       &  \\
    $Y_4$   & 0.0056358 &       &       &  \\
    \textbf{} & \multicolumn{4}{c}{\textbf{Fast neutron detections in Region B}} \\
    $Y_1$    & -0.0000243761 & -0.000134373 & 0.000823734 & 0.0126651\\
    $Y_2$    &-0.000094874 & 0.000795215 & 0.0544565 &  \\
    $Y_3$    & 0.000301926 &       &       &  \\
    $Y_4$   &0.000161625&       &       &  \\
    \end{tabular}%
  \label{table:tab1}%
\end{table}%

\begin{table}[htbp]\footnotesize
\caption{The values of the $Y_i$ and $\omega_i$ calculated for four various versions of Feynman-alpha formulas (the source is in Region B, fast neutron detector is used either in Region A or B).}
  \centering
    \begin{tabular}{rrrrr}
        \textbf{} & \multicolumn{4}{c}{\textbf{Source in Region B}} \\
          & \textbf{2-point} & \textbf{2-point } & \textbf{1-point} & \textbf{1-point} \\
    \textbf{} & \textbf{2-group} & \textbf{1-group} & \textbf{2-group} & \textbf{1-group} \\
    $\omega_1$    & 1.00891 & 1.06295 & 1.12123  & 0.905974 \\
    $\omega_2$   &0.721378 & 0.0660208 & 0.516779 &  \\
    $\omega_3$    & 0.28402 &       &       &  \\
    $\omega_4$    & 0.00211384 &       &       &  \\
    \textbf{} & \multicolumn{4}{c}{\textbf{Fast neutron detections in Region A}} \\
    $Y_1$    & 5.90179E-7 & 1.29166E-6& 9.406E-6 & 0.943027E-4\\
     $Y_2$     & 4.20869E-6 &0.014281 & 0.122662E-3 &  \\
     $Y_3$     & 0.159498E-2&       &       &  \\
     $Y_4$     & 1.07716 &       &       &  \\
    \textbf{} & \multicolumn{4}{c}{\textbf{Fast neutron detections in Region B}} \\
     $Y_1$     & -2.65992E-9 & -2.18905E-7 &  9.406E-6 & 0.943027E-4 \\
     $Y_2$     & -1.33593E-6 & 0.567443E-4& 0.122662E-3&  \\
    $Y_3$     &8.38725E-6&       &       &  \\
     $Y_4$     & 0.00477401&       &       &  \\
    \end{tabular}%
\label{table:tab2}
\end{table}%

\begin{table}[htbp]\footnotesize
\caption{The values of the $Y_i$ and $\omega_i$ calculated for four various versions of Feynman-alpha formulas (the source is in Region A, thermal neutron detector is used either in Region A or B).}
  \centering
    \begin{tabular}{rrrrr}
        \textbf{} & \multicolumn{4}{c}{\textbf{Source in Region A}} \\
          & \textbf{2-point} & \textbf{2-point } & \textbf{1-point} & \textbf{1-point} \\
    \textbf{} & \textbf{2-group} & \textbf{1-group} & \textbf{2-group} & \textbf{1-group} \\
     $\omega_1$    & 1.52001 &1.8083 & 1.45471 & 0.923611 \\
    $\omega_2$     & 1.12141 &0.743335 & 0.350087 &  \\
    $\omega_3$    & 0.759289 &       &       &  \\
    $\omega_4$     & 0.0983484&       &       &  \\
    \textbf{} & \multicolumn{4}{c}{\textbf{Thermal neutron detections in Region A}} \\
    $Y_1$    & -1.31014E-7& 0.00154341 & -0.806821E-4 & 0.0126651 \\
    $Y_2$    & 9.33539E-7 &0.00938149 & 0.00139308 &  \\
    $Y_3$    &-2.39735E-6 &       &       &  \\
    $Y_4$    &0.000052813 &       &       &  \\
    \textbf{} & \multicolumn{4}{c}{\textbf{Thermal neutron detections in Region B}} \\
    $Y_1$    & 0.108441E-3 &-0.134373E-3 & -0.806821E-4 & 0.0126651\\
    $Y_2$    &-0.505571E-3&0.795215E-3 & 0.00139308 &  \\
    $Y_3$    &0.663882E-3 &       &       &  \\
    $Y_4$    &0.258841E-3&       &       &  \\
    \end{tabular}%
  \label{table:tab3}
\end{table}%

\begin{table}[htbp]\footnotesize
\caption{The values of the $Y_i$ and $\omega_i$ calculated for four various versions of Feynman-alpha formulas (the source is in Region B, thermal neutron detector is used either in Region A or B).}
  \centering
    \begin{tabular}{rrrrr}
        \textbf{} & \multicolumn{4}{c}{\textbf{Source in Region B}} \\
          & \textbf{2-point} & \textbf{2-point } & \textbf{1-point} & \textbf{1-point} \\
    \textbf{} & \textbf{2-group} & \textbf{1-group} & \textbf{2-group} & \textbf{1-group} \\
    $\omega_1$    & 1.00891 & 1.06295 & 1.12123  & 0.905974 \\
    $\omega_2$   &0.721378 & 0.0660208 & 0.516779 &  \\
    $\omega_3$    & 0.28402 &       &       &  \\
    $\omega_4$    & 0.00211384 &       &       &  \\
    \textbf{} & \multicolumn{4}{c}{\textbf{Thermal neutron detections in Region A}} \\
    $Y_1$    &-8.84973E-9 & 1.29166E-6& -4.82586E-6 & 0.943027E-4\\
    $Y_2$   & 4.87895E-8& 0.014281 &0.227173E-4 &  \\
    $Y_3$    &-5.6035E-6&       &       &  \\
    $Y_4$    &0.0974942 &       &       &  \\
    \textbf{} & \multicolumn{4}{c}{\textbf{Thermal neutron detections in Region B}} \\
    $Y_1$    &1.71005E-6 & -2.18905E-7 &-4.82586E-6 & 0.943027E-4\\
    $Y_2$    &-7.10267E-6 & 0.567443E-4 & 0.227173E-4&  \\
    $Y_3$    &0.232948E-4 &       &       &  \\
    $Y_4$    &0.0170835&       &       &  \\
    \end{tabular}
  \label{table:tab4}
\end{table}%

Since there is only one region considered in the two-group one-point and the one-point one-group Feynman-alpha formulas, the coefficients are the same for the detection in the different regions of the initial system used for the simulations. The same is true for the energy-dependent factor in the two-point one-group and the one-point one-group Feynman-alpha formulas, the coefficients are the same for the fast and thermal neutron detection.

\textbf{Attention!} In the studies described below we assume that the two-point two-group version of Feynman-alpha formulas gives the most accurate predictions as the most involved one among the four various versions, i.e. the two-point two-group, the two-point one-group, the two-group one-point and one-point one-group theories.
\subsection{Comparison of the four versions of the Feynman-alpha theoretical formulas for the case of fast neutrons detections}
      \label{sec:12}
\begin{figure}[ht!]
\centering
  \includegraphics[width=0.99\linewidth]{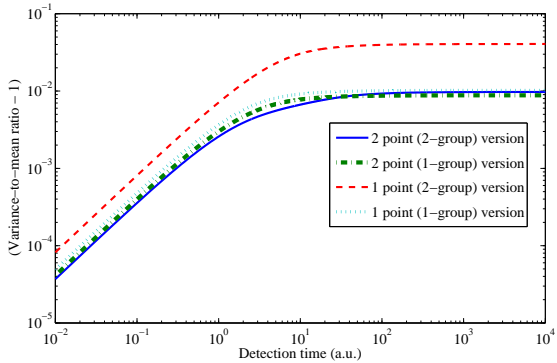}
\caption{The dependence of the ratio of the variance to mean of the number of fast neutron detections on the detection time for four versions of Feynman-alpha theory (the source is in Region A, detector is in Region A).}
\label{fig:4}       
\end{figure}

\begin{figure}[ht!]
\centering
  \includegraphics[width=0.99\linewidth]{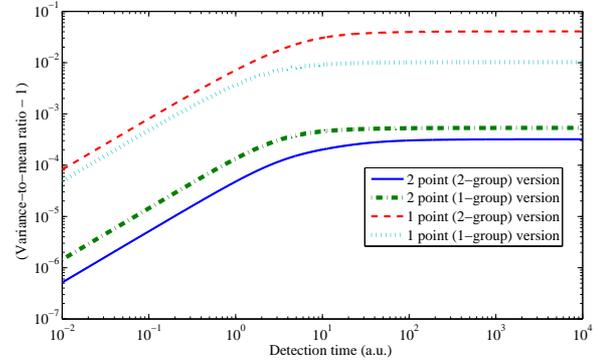}
\caption{The dependence of the ratio of the variance to mean of the number of fast neutron detections on the detection time for four versions of Feynman-alpha theory (the source is in Region A, detector is in Region B).}
\label{fig:5}       
\end{figure}

Figures \ref{fig:4}-\ref{fig:5} show a quantitative illustration of the dependence of the variance to mean of the number of fast neutron detections on the detection time for four versions of Feynman-alpha theories when the source is in Region A. Different curves in Figures \ref{fig:4}-\ref{fig:5} are created based on the parameter values from Table \ref{table:tab1}-\ref{table:tab2}.

As it is shown in Figure \ref{fig:4}, when fast neutrons are detected in Region A, the two-point two-group, the two-point one-group and one-point one-group versions of the Feynman-alpha theoretical formulas give very similar results. However, the one-point two-group version of the formulas overestimates the asymptotic ratio of the variance to mean. Thus, we can conclude that the region dependence of the model plays a more important role than the energy dependence for the case when the source and the fast neutron detector are both placed in the region of the nuclear material. Therefore, in this situation all three versions of the Feynman-alpha theory, the two-point two-group, the two-point one-group and one-point one-group, can be used, although it is more time-efficient to use the one-group one-point version of the Feynman-alpha theory. As an example, in reality this case may be related to the measurements performed in the spent fuel pool when the detector is placed in the control tube of fuel assembly.

In the case when the fast neutron detector is in Region B (Figure \ref{fig:5}), a slight difference is observed between the two-point two-group and the two-point one-group versions of Feynman-alpha theories. At the same time, the one-point two-group and one-point one-group versions of Feynman-alpha theory significantly overestimate the values of variance to mean ratio obtained with the two-point two-group version of the formulas. Thus, in this case two versions of Feynman-alpha theory, the two-point two-group and the two-point one-group can be used, although it is more time-efficient to use the two-point one-group version for quantitative estimates.

\begin{figure}[ht!]
\centering
  \includegraphics[width=0.99\linewidth]{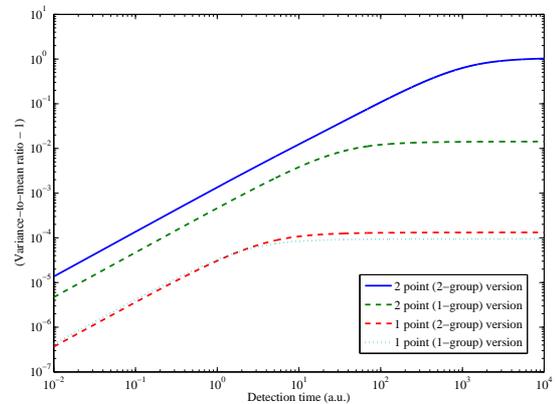}
\caption{The dependence of the ratio of the variance to mean of the number of fast neutron detections on the detection time for four versions of Feynman-alpha theory (the source is in Region B, detector is in Region A).}
\label{fig:6}       
\end{figure}

\begin{figure}[ht!]
\centering
  \includegraphics[width=0.99\linewidth]{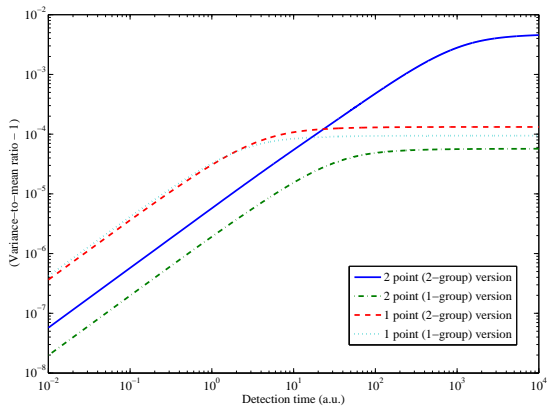}
\caption{Dependence of the ratio of the variance to mean of the number of fast neutron detections on the detection time for four versions of Feynman-alpha theory (the source is in Region B, detector is in Region B).}
\label{fig:7}       
\end{figure}

The differences between the various versions of Feynman-alpha theory are significantly higher when the neutron source is placed in Region B and the fast neutron detector is in either Region A or B, see Figures \ref{fig:6}-\ref{fig:7}.

In all cases when the fast neutron detector is used, the two-point two-group version of Feynman-alpha formulas produce high values of the asymptotic variance-to-mean ratio compared to results obtained with other versions, i.e. the two-point one-group, one-point two-group and one-point one-group versions of the theory.
      \subsection{Comparison of four versions of Feynman-alpha theoretical formulas for the case of thermal neutron detections}
\label{sec:13}
\begin{figure}[ht!]
\centering
  \includegraphics[width=0.99\linewidth]{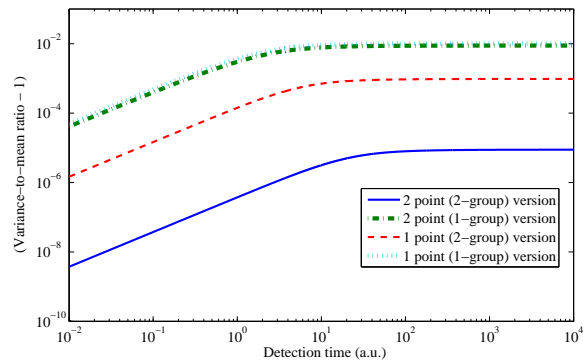}
\caption{The dependence of the ratio of the variance to mean of the number of thermal neutron detections on the detection time for four versions of Feynman-alpha theory (the source is in Region A, detector is in Region A).}
\label{fig:8}       
\end{figure}

Regarding thermal neutron detection, when the source and the detector are in Region A (Figure \ref{fig:8}), the three special versions of the Feynman-alpha theory, the two-point one-group, the one-point two-group and the one-point one-group, all deviate significantly from the two-point two-group version. However, the one-point one-group theory gives very similar predictions of the ratio of the variance to mean as the two-point one-group theory. At the same time, the two-group one-point theory provides somewhat more accurate results. Thus, the impact of the energy-dependence appears to be somewhat higher than the impact of the space-dependence.

\begin{figure}[ht!]
\centering
  \includegraphics[width=0.99\linewidth]{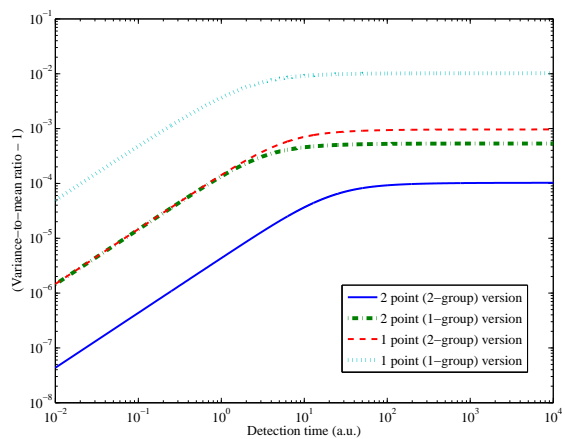}
\caption{The dependence of the ratio of the variance to mean of the number of thermal neutron detections on the detection time for four versions of Feynman-alpha theory (the source is in Region A, detector is in Region B).}
\label{fig:9}       
\end{figure}

When the thermal neutron detection is performed in Region B (Figure \ref{fig:9}), we may conclude that both the space-dependent and energy-dependent aspects play important role for this case.

\begin{figure}[ht!]
\centering
  \includegraphics[width=0.99\linewidth]{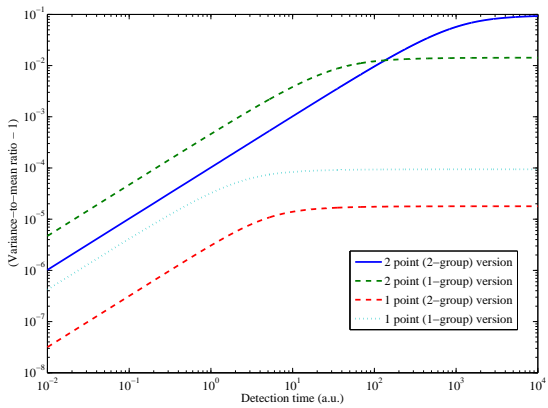}
\caption{Dependence of the ratio of the variance to mean of the number of thermal neutron detections on the detection time for four versions of Feynman-alpha theory (the source is in Region B, detector is in Region A). }
\label{fig:10}       
\end{figure}

If the source is in Region B and detection is performed in Region A (Figure \ref{fig:10}), the two-point one-group version of the Feynman-alpha theory gives results which are closer to the one obtained with two-point two-group theory. Thus, the impact of the space-dependence to the final results is higher than the impact of energy-dependence.

\begin{figure}[ht!]
\centering
  \includegraphics[width=0.99\linewidth]{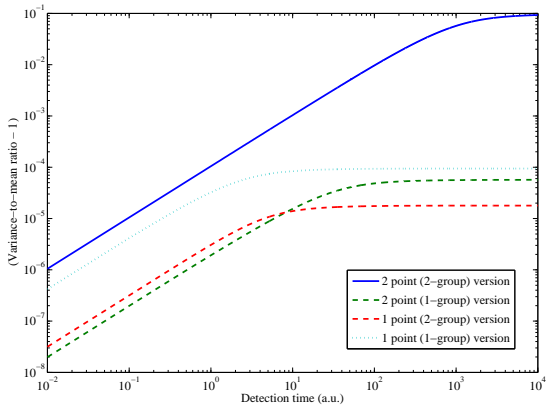}
\caption{Dependence of the ratio of the variance to mean of the number of thermal neutron detections on the detection time for four versions of Feynman-alpha theory (the source is in Region B, detector is in Region B). }
\label{fig:11}       
\end{figure}

Although, for a case of detector and source being placed in Region B (Figures \ref{fig:11}), the two-point one-group, two-group one-point and  one-group one-point versions provide results of the variance-to-mean ratio that are significantly deviating from the ratio obtained by using the two-point two-group theory. Thus, the space-dependent and energy-dependent aspects, both play the important role in this situation.

In general, we can say that for the thermal neutron detections when the detection is done in Region A, the energy-dependence has a higher impact to the ratio of the variance to mean than the space-dependent factor. On the other hand, for detection in Region B both factors should be equally taken into account.
\section{Conclusions}
      \label{sec:14}
The two-group two-point version of Feynman-alpha theory was derived with a use of the forward master equation technique. The two-group one-point Feynman-alpha theory (with delay neutrons) is extended by including fast neutron detections and and fast fissions. The two-point one-group variance-to-mean formula (with delay neutrons) is enhanced as well, by including detection and source terms in both regions. Thus, this gives the possibility of treating fast reflected systems in a more accurate way, by treating the counts separately in the fast and the thermal groups (or in the nuclear material (fissile region) and reflector regions).

The comparative study of the two-group two-point, the two-group one-point, the one-group two-point and the one-group one-point Feynman-alpha models is made by using the specific reaction intensities obtained in Monte-Carlo simulations. It is shown that for all cases when the fast neutron detector is used in measurements, the space-dependent aspect has a higher impact on the final results than the energy-dependent aspect. In particular, when the source and the fast neutron detector, both placed in the region of nuclear material three versions of Feynman-alpha theory provides similar accuracy in the determination of the variance to mean ratio. Namely, the two-point two-group, the two-point one-group and one-point one-group can be used, although it is more time-efficient to use the one-group one-point version of the Feynman-alpha theory. The situation is not so optimistic for a case, when the fast neutron detector is in Region B, because a slight difference is observed between all versions of the theories. The one-point two-group and one-point one-group versions of Feynman-alpha theory significantly overestimate the values of variance to mean ratio obtained with the two-point two-group version of the formulas. Therefore, in this case two versions of Feynman-alpha theory, the two-point two-group, the two-point one-group can be considered as the accurate qualitative estimates. Regarding the use of the thermal neutron detections, both energy- and space-dependent factors are important to take into account. However, when the detection is done in Region A, the space-dependence has a higher impact to the ratio of the variance to mean than the energy-dependence, while, for detection in Region B both factors should be equally considered.

\begin{acknowledgements}
This work was supported by the  \textbf{Swedish Radiation Safety Authority, SSM}. The authors want to thank \textbf{Dr. Stephen Croft} for useful discussions and advice.
\end{acknowledgements}



\begin{thebibliography}{}
%
\bibitem{Fermi} {Fermi, E., Feynman, R. P., de Hoffmann, F., Theory of the Criticality of the Water Boiler and the Determination of the Number of Delayed Neutrons, \textit{USAEC Report
MDDC-383 (LADC-269)}, Los Alamos Scientific Laboratory, 1944.}
\bibitem{Hoffmann} {de Hoffmann, F., Intensity Fluctuations of a Neutron Chain Reaction, \textit{USAEC Report MDDC-382 (LADC-256)}, Los Alamos Scientific Laboratory, 1946.}
\bibitem{Feynman} {R. P. Feynman, F. de Hoffmann, and R. Serber, Journal of Nuclear Energy 64 (1956).}
\bibitem{Imre} {I. P\'{a}zsit and L. P\'{a}l, Neutron Fluctuations: A Treatise on the Physics of Branching Processes. (Elsevier Science Ltd., London, New York, Tokyo, 2008).}
\bibitem{Kiyavitskaya} {H. Kiyavitskaya and e. al., YALINA-Booster Benchmark Specifications for the IAEA Coordinated Research Projects on Analytical and Experimental Benchmark Analysis on Accelerator Driven Systems and Low Enriched Uranium Fuel Utilization in Accelerator Driven Sub-Critical Assembly Systems, edited by IAEA (IAEA, Vienna, 2007).}
\bibitem{Gohar}{Y. Gohar and D. L. Smith, \textit{Report YALINA Facility, A Sub-Critical Accelerator-Driven System (ADS) for Nuclear Energy Research. Facility Description and Overview of the Research Program}, Argonne National Laboratory, ANL-10/05, 2010.}
\bibitem{Talamo}{A. Talamo and Y. Gohar, \textit{Deterministic and Monte Carlo Modeling of YALINA Thermal Subcritical Assembly}, Argonne National Laboratory, ANL-10/17, 2010.}
\bibitem{Soule} {R. Soule, W. Assal, P. Chaussonnet, and e. al., Nuclear Science and Engineering 148 (1), 124 (2004).}
\bibitem{Munoz} {J.-L. Munoz-Cobo, C. Berglof, J. Pena, D. Villamarin, and V. Bournos, Annals of Nuclear Energy 38 (2011).}
\bibitem{Berglof} {C. Berglof, M. Fernandez-Ordonez, D. Villamarin, and et. al., Annals of Nuclear Energy 38 (2011).}
\bibitem{Rana2009}{Y. S. Rana and S. B. Degweker, Nuclear Science and Engineering 162 (2), 117 (2009).}
\bibitem{Rana2011}{Y. S. Rana and S. B. Degweker, Nuclear Science and Engineering 169, 98 (2011).}
\bibitem{Degweker2000}{S. B. Degweker, Annals of Nuclear Energy 30, 223 (2000).}
\bibitem{Yamane2000}{Yamane, Y., Uritani, A., Ishitani, K., Kataoka, H., Shiroya, S., Ichihara, C., Kobayashi, K., Okajima, S., 2000. Experimental study on reactor dynamics of accelerator driven subcritical system. \textit{KURRI Progress Report}, vol. 46. Research Reactor Institute, Kyoto University, 1999.}
\bibitem{Yamane2001}{Yamane, Y., Uritani, A., Kataoka, H., Nomura, T., Matsui, M., Naka, R., Tabuchi, A., Hayashi, T., Yamauchi, H., Kitamura, Y., Shiroya, S., Misawa, T., Kobayashi, K., Ichihara, C., Unesaki, H., Nakamura, H., Experimental study on reactor dynamics of accelerator driven subcritical system (II). \textit{KURRI Progress Report}, vol. 53. Research Reactor Institute, Kyoto University, 2001.}
\bibitem{Yamane2002}{Y. Yamane, Y. Kitamura, H. Kataoka, K. Ishitani, and S. Shiroya, in Proceedings of PHYSOR-2002 International Conference on the New Frontiers of Nuclear Technology: Reactor Physics, Safety and High-Performance Computing (Seoul, Korea, 2002).}
\bibitem{Kitamura2003}{Y. Kitamura, H. Yamauchi, and Y. Yamane, Annals of Nuclear Energy 30, 897 (2003).}
\bibitem{Kitamura2004}{Y. Kitamura, H. Yamauchi, Y. Yamane, T. Misawa, C. Ichihara, and H. Nakamura, Annals of Nuclear Energy 31, 163 (2004).}
\bibitem{Kitamura20044}{Y. Kitamura, K. Taguchi, A. Yamamoto, Y. Yamane, T. Misawa, C. Ichihara, H. Nakamura, and H. Oigawa, International Journal of Nuclear Energy Science and Technology (2004).}
\bibitem{Kitamura2005}{Y. Kitamura, I. P\'{a}zsit, J. Wright, A. Yamamoto, and Y. Yamane, Annals of Nuclear Energy (2005).}
\bibitem{Munoz2001}{J. L. Munoz-Cobo, Y. Rugama, T. E. Valentine, J. T. Michalzo, and R. B. Perez, Annals of Nuclear Energy 28, 1519 (2001).}
\bibitem{Degweker2003}{S. B. Degweker, Annals of Nuclear Energy 30, 223 (2003).}
\bibitem{Ceder2003}{M. Ceder and I. P\'{a}zsit, Progress in Nuclear Energy 43, 429 (2003).}
\bibitem{Imre2004} {I. P\'{a}zsit, M. Ceder, and Z. Kuang, Nuclear Science and Engineering 148, 67 (2004).}
\bibitem{Croft} {S. Croft, A. Favalli, D.K. Hauck, D. Henzlova, P.A. Santi, Nuclear Instruments and Methods A 686, 136 (2012).}
\bibitem{Menlove} {H. O. Menlove, S. H. Menlove, and S. T. Tobin, Nuclear Instruments and Methods A 602, 588 (2009).}
\bibitem{Anderson2012} {J. Anderson, L. P\'{a}l, I. P\'{a}zsit, D. Chernikova, and S. Pozzi, The European Physical Journal Plus 127, 21 (2012).}
\bibitem{Pal2012}{L. P\'{a}l and I. P\'{a}zsit, Sci. Technol. Nucl. Install. (2012).}
\bibitem{Yamamoto2013}{T. Yamamoto, Annals of Nuclear Energy 57, 84 (2013).}
\bibitem{Chapline2011}{G. Chapline, in Proceedings of 15th International conference on Emerging nuclear Energy systems (San Francisco, 2011).}
\bibitem{Anderson20121}{J. Anderson, D. Chernikova, I. P\'{a}zsit, L. P\'{a}l, and S. A. Pozzi, The European Physical Journal Plus 127, 90 (2012).}
\bibitem{Chernikova2013} {D. Chernikova, I. P\'{a}zsit, and W. Ziguan, in Proceedings of ESARDA meeting (2013).}
\bibitem{Chernikova20131} {D. Chernikova, I. P\'{a}zsit, L. P\'{a}l and W. Ziguan, in Proceedings of INMM meeting (2013).}
\end{thebibliography}
\end{document}